\documentclass{aa}

\newcommand{\msol}{$M_{\odot}$}
\newcommand{\msolyr}{$M_{\odot}$\,yr$^{-1}$}
\newcommand{\hi}{H\,{\small{\sc I}}}

\newcommand{\kms}{km s$^{-1}$}

\usepackage{graphicx}
\usepackage{subfigure}
\usepackage{afterpage}

\begin{document}

\hyphenation{con-sti-tu-ents there-by mar-gin-al-ly stat-is-tics}

%\thesaurus{03(11.08.1, 11.09.1: NGC\,1406, NGC\,1511, NGC\,4244,
%  NGC\,4565, NGC\,4700, NGC\,7090, NGC\,7462, 11.19.3, 13.18.1)}

\title{Optical and radio survey of Southern Compact Groups of galaxies. I.
Pilot study of six groups.}

\authorrunning{E. Pompei et al.}
\titlerunning{Optical and radio study of six southern compact groups}

\author{E. Pompei\inst{1}
\and
M. Dahlem\inst{2}
\and
A. Iovino\inst{3}
}

\institute{European Southern Observatory (ESO), Alonso de Cordova 3107, Santiago,
Chile
\and
CSIRO/ATNF Paul Wild Observatory, Locked Bag 194,
Narrabri NSW 2390, Australia
\and
Osservatorio Astronomico di Brera, Via Brera 28, Milan, Italy
}

\offprints{E.P., epompei@eso.org}

\date{Received xx mmm 2006? / Accepted 19 07 2007}

\abstract{Multi-wavelength observations of Hickson's Compact Groups (HCGs) have shown
that many of these groups are physical bound structures and are in different stage
of evolution, from spiral-dominated systems to almost merged objects. Very few studies
have analysed the Southern Compact Groups (SCGs) sample, which is though to be
younger that HCGs, due to an on average higher number of spiral galaxies. We present here
the first results from optical and radio observations on a pilot sample of SCGs.}
{Previous HI studies of HCGs seems to corroborate an evolutionary sequence from
low velocity dispersion, spiral dominated, young and star-forming groups to higher
velocity dispersion, mostly early-type, X-ray bright groups.
In this scheme, the dominant parameter is most likely the ratio between the galaxy
mass vs the available gas mass. A second parameter, namely the merging history
of the group, however might have an important role in the evolution of the
intra-group medium. Powerful merging events and/or interactions can contribute
significantly to the heating of the intra-group medium by means of supernova explosions or by 
triggering an AGN. We propose to investigate whether the evolutionary scheme of the
intragroup medium found for HCGs also holds in a different compact group sample. In
addition to this, we start in this paper to investigate which is the
influence of the merging history of groups on their evolution.}
{Optical observations of SCGs obtained with ESO telescopes and radio data from
the ATCA allow us to probe
the distribution of the warm intra-group medium and the evolutionary stage of each
group, by means of morphological studies and via measurements of star formation and other
types of nuclear activity.}
{We present here results from a pilot sample of SCGs: we find an evolutionary trend, based
on the properties of the warm intra-group medium, similar to what has been found for HCGs. 
Both galaxies with Sy2 nuclei in our sample are members of groups in the late stages of their
dynamical evolution. However while one group is strongly HI deficient and shows, at the same
time a strong ongoing merging, the other does not show any HI deficiency and the galaxies
are only interacting with each other.}
{}

\maketitle

\section{Introduction}
Compact groups of galaxies are among the densest and smallest associations
of galaxies which can be found on the sky. Comprising 4 to 10 galaxies, 
with a galaxy-galaxy separation comparable to the
individual galaxy diameters and a velocity dispersion of the order
of 200 \kms, they are the best-suited laboratories for studies
of strong galaxy interactions; even more so in the absence of massive elliptical
galaxies and their associated gravitationally heated X-ray-emitting
intergalactic envelopes, as found in massive clusters.

However strong mergings in the Hickson's
compact group sample are rare, of the order of 6-7$\%$ of the total
number of galaxies (Mendes de Oliveira et al., 1994).
Tidal interactions, on the other hand, occur frequently, with 50$\%$
of the galaxies in Hickson's sample showing signs of morphological
disturbances. 

The interactions often involve more than
two galaxies, which leads to a large number of targets for
investigations of the effects of tidal forces on the member
galaxies. 

In such close proximity it is also likely that
tidal interactions, galaxy harassment and similar processes
play a major role in the evolution of the galaxies and of their
morphological type. 
Hickson's groups show a remarkable lack of spiral galaxies with
respect to field galaxies, 49$\%$ against 82$\%$, and approximately
2/3 of the groups show diffuse X-ray intra-group emission (Ponman et al., 1996).
While some groups, like HCG 16 (Ribeiro et al, 1998), are still very active
and rich in spirals, most of the HCGs appear to be in an advanced stage
of evolution, rich in early type galaxies and surrounded by an hot intragroup medium.
The warm intragroup medium seems to follow an evolutionary
sequence driven mainly by the ratio of galaxy mass vs gas mass (Verdes-Montenegro
et al., 2001, herafter VM01): from \hi\ mainly centered on member galaxies to groups
without any warm gas at all or to groups with diffuse \hi\, often offset 
with respect to the stellar components of the member galaxies.

The motivation for the work presented here is to investigate how
groups richer in late type galaxies will evolve and what is the fate
of their warm gas.\newline
The Southern Compact Group sample (hereafter SCGs), 
described in detail by Iovino (2002, hereafter I02),
is richer in spirals than HCGs. In fact, $\sim$ 70$\%$ of its member galaxies
are spirals.

The catalog, based on the COSMOS scans of the SERC(J) southern sky survey, 
has been created following criteria very similar
to the Hickson's ones, but allowing for faint galaxies in the so
called {\it isolation ring}, thus avoiding one of the biases of Hickson's
catalog, i.e. to discard groups whose fainter galaxy members had a magnitude
close to the cut-off limit. For more details, we refer the reader to
the cited work, however, for completeness sake, we report here
the selection criteria:
\begin{itemize}
\item{} Richness: n$\ge$ 4, where n is the number of galaxies
within $\Delta$mag$_{comp}$ $\le$ 3, where $\Delta$mag$_{comp}$ is the
magnitude difference between the brightest group member (m$_{brightest}$)
and the faintest (m$_{faintest}$). 
\item{} Isolation: R$_{isolation} \ge$ 3 $\times$ R$_{group}$ where 
R$_{group}$ is the radius of the smallest circle containing the centres of
all group members and R$_{isolation}$ is the distance from the centre of
the circle to the nearest non-member galaxy within 0.35 mag from the faintest
group member.
\item{} Compactness: $\mu_{gr} \le \mu_{limit}$, where $\mu_{gr}$ is
the mean surface brightness within the circle of radius  R$_{group}$
and $\mu_{limit}$ = 27.7 in the b$_{j}$ band.
\end{itemize}

A bright subsample of the SCGs, i.e. all those groups whose
brightest galaxy has b$_{j}$ $\le$ 14.5, for a total of 50 groups, has been extensively
observed in the optical: results on the star formation history and galaxy activity in SCGs
have been presented by Coziol et al. (2000). 
At the moment, however, very little information is available 
on the SCGs intragroup medium, either warm or hot: this means that we
lack an important diagnostic of the evolutionary history of the groups and of
the interactions among member galaxies.

To this aim, we have started a multi-wavelength observing campaign, including
optical imaging and spectroscopy, as well as radio
observations. 
Here we present, as a first showcase, optical, \hi\, and radio
continuum data of a small number of objects, for which good-quality
data are available already.

Optical imaging in the R filter traces the stellar distributions and provides
information on morphology; spectroscopy, through measurements
of abundances and star formation rates,  will tell us whether galaxies
have predominantly old or young stellar populations. 

\hi\ is the best tracer of gas being pulled out of galaxies
taking part in gravitational interactions.

Radio continuum and H$\alpha$ trace star formation (SF) triggered by
radial flows in the disturbed gravitational potential of the
group members.

Star formation in three SCGs, including SCG\,0018-4854, is described
briefly by Temporin et al. (2005).

This paper is divided as follows: in Sect. 2 we describe the optical and radio
observations and data reduction, in Sect. 3 we present our results, which
are discussed in Section 4. Sect.5 contains our conclusions.

Throughout the paper, a $\Lambda$CDM cosmology ($\Omega_{M}$=0.3;
$\Omega_{\Lambda}$=0.7) and H$_{0}$=67 \kms Mpc$^{-1}$ have
been used.

\section{Observations and data reduction}
The targets studied by us were selected from the brighter subsample of the Southern
Compact Groups Survey (I02), based on target visibility during the allocated
observing periods. All targets, with the exception of SCG2159-3210, 
are spiral dominated groups, and they represent
a fair pilot sample of the SCGs survey.
Their main characteristics are given in Table 1.

\begin{table*}
\caption{Salient parameters of six Southern Compact Groups of galaxies. $N_{\rm conc}$ is the number of 
concordant galaxies in the original SCGs catalog, while $N_{\rm Gal}$ is the total number of confirmed member
galaxies fulfilling the group selection criteria and having a similar redshift from optical and radio data.}
\label{groups}
\begin{minipage}{\textwidth}
\begin{tabular}{lcccccc}
\hline
Group  name     &  Alternate name & Centre Position\footnote{Approximate centre positions; pointing centres of radio observations}
 & $v_{\rm hel}$\footnote{Mean value of all measured optical recession velocities}  &
 D\footnote{Based on $H_\circ$ = 67 km s$^{-1}$ Mpc$^{-1}$, $\Omega_{\rm m}=0.3$ and $\Omega_\lambda=0.7$} & $N_{\rm conc}$ & $N_{\rm Gal}$ \\
                &                 & $\alpha,\delta$(2000)          & (km s$^{-1}$)                & (Mpc)      &                   &      \\ \hline
SCG\,2159-3210  &    HCG90        & 22:02:02.6, --31:55:26         & 2643                        & 35.5        & 4            & 4 \\
SCG\,2353-6101  &     ---         & 23:55:47.0, --60:43:06         & 4538                        & 61.2        & 3            & 3 \\
SCG\,0018-4854  &    AM0018-485   & 00:21:23.1, --48:38:19         & 3386                        & 45.5        & 4            & 5 \\
SCG\,0122-3819  &     ---         & 01:25:16.0, --38:08:35         & 6125                       & 82.9        & 4            & 5 \\
SCG\,0141-3429  &    USCG S063    & 01:43:24.0, --34:14:42         & 3905                       & 52.6        & 5            & 5 \\
SCG\,0227-4212  &     ---         & 02:29:30.9, --43:01:41         & 5384                      & 72.8        & 6            & 6 \\\hline
\end{tabular}
\end{minipage}
\end{table*}

\subsection{Optical data}

Optical images and spectra of the groups presented
here have been obtained at La Silla Observatory during
different runs. We discuss below the main results on six groups obtained from
these data, while we defer the full presentation of the
optical data to a future paper.

\subsubsection{Optical broad-band imaging}

Optical images of a large number of SCGs have been obtained at
the Danish 1.54m telescope, with the Danish Faint Object Spectrograph and Camera
(DFOSC) in the R filter during two separate runs, in August 1995 and August 1996.
The data presented in this paper are part of the 1996 run only; during this year the DFOSC
CCD was upgraded to a 2048 $\times$ 2048 pixels array, 
with a pixel size of 0.39$\arcsec$/pix and a field of view of 14$\arcmin$ $\times$ 14$\arcmin$;
the seeing during the observations varied between 1.0$\arcsec$ and 1.4$\arcsec$.
In this paper we present also additional data, again with the R filter, acquired in January 2006 
using Eso Faint Object Spectrograph
and Camera 2 (EFOSC2) on the 3.6m telescope. The 2048$\times$ 2048 pixels CCD has a field of
view of 5.4$\arcmin$ $\times$ 5.4$\arcmin$ and a pixel scale of 0.157$\arcsec$/pixel and it
was used in binned mode; average seeing on the images was 0.9$\arcsec$.

Data reduction has been carried out with IRAF\footnote{Image Reduction and 
Analysis Facility, written and supported by the IRAF programming group at 
the National Optical Astronomy Observatories (NOAO)} in the standard way;
photometric zero points have been measured from Landolt fields observed
during the same nights of the observations. A log
of the observations and the zero-points is shown in Table 2.

\setcounter{table}{1}
\begin{table*}
\caption{Observing log; SCG\,2159-3210 has not been observed by us in the optical and related results are taken from the literature.}
\label{optobs}
\begin{tabular}{ccccccc}
\hline
Group           & Date of observations & Telescope and instrument & Filter & Pixel scale & Exposure time & Zero-Point \\
                &              &                       &              & (arcsec)    & (seconds) &             \\\hline
SCG\,2353-6101  & August 1996  & Danish 1.54m+DFOSC    & ESO\#452 (R) &   0.39      &  3 x 200  & 23.92$\pm$ 0.03 \\
SCG\,0018-4854  & August 1996  & Danish 1.54m+DFOSC    & ESO\#452 (R) &   0.39      &  3 x 200  & 23.92$\pm$ 0.03 \\
SCG\,0122-3819  & August 1996  & Danish 1.54m+DFOSC    & ESO\#452 (R) &   0.39      &  3 x 200  & 23.92$\pm$ 0.03 \\
SCG\,0141-3429  & January 2006 & 3.6m telescope+EFOSC2 & ESO\#642 (R) &   0.32      &  5 x 90   & 26.20$\pm$ 0.02 \\
SCG\,0227-4212  & August 1996  & Danish 1.54m+DFOSC    & ESO\#452 (R) &   0.39      &  3 x 200  & 23.92$\pm$ 0.03 \\ \hline
\end{tabular}
\end{table*}

The GALFIT software (Peng et al., 2002) was used to fit the surface brightness
distribution of the member galaxies of each group and identifying the
different components (bulge, bar, disk and central source) and their
respective brightness and scale length. The isophotes could be reliably traced down 
to 25.4 magnitude in R, and results for all galaxies are shown in Table 3 at the end of the paper.
Postage-stamp
images of the galaxies, the models obtained from the software and the
residuals are shown in Figure 1.
Residual maps from the fitting are used to identify
small scale structures, which would otherwise have remained undetected on
the original images.

\subsubsection{Optical spectroscopy}

The complete brighter subsample of the SCGs has been observed at the
ESO 1.52m telescope equipped with the Boller\&Chivens spectrograph
in several runs from 1995 to 1997. Instrument set-up, data reduction and analysis,
as well as the complete spectroscopic classification of the
groups, with the exception of SCG0141-3429, can be found in Coziol et al. (2000).
 
More sensitive spectroscopic observations have been obtained at the
3.6m telescope with the EFOSC2 spectro-imager between
1999 and 2006, including also fainter groups from the SCGs catalog.
The increased depth of these data and the
improved spatial sampling allow for a more precise
spectroscopic classification.\newline
More detailed spectroscopic results for studies of the chemical evolution of galaxies and star
formation history in compact
groups will be presented elsewhere, while here 
we use optical spectroscopic results to
measure recession velocities, and thereby establish group
membership of galaxies, and to classify which type of nuclear
activity is observed in the member galaxies of each group.

Observations of SCG0141-3429 are presented here for the first time;
this is due to the fact that the magnitude of its brightest
galaxy is 14.54, which led to discard this group from the brightest SCGs
subsample.
The data were obtained in 2006 with the 3.6m telescope and EFOSC2, 
equipped with grism \#11 and 1$\arcsec$ slit, reading the
CCD in binned mode. Standard
data reduction was carried on using MIDAS\footnote{Munich Image Data Analysis
System, developed and maintained by European Southern Observatory (ESO).} and a dedicated set of scripts; unfortunately
the conditions of the night were not optimal (thin clouds and airmass ranging from 1.2 to 1.9), and
no flux calibration was possible.
The spectra however are of sufficient quality to allow a preliminary
classification of the galaxies activity based on line ratios.

\subsection{ATCA radio data}

Data of six SCGs were obtained with the Australia Telescope
Compact Array (ATCA\footnote{The Australia Telescope is funded
by the Commonwealth of Australia for operation as a National
Facility managed by CSIRO.}) in 1995, a short summary of
which was given by Oosterloo and Iovino (1997). We have now
retrieved the data from the public archive and reprocessed them.
In one case, SCG\,0122-3819, more data were obtained in the
meantime to improve the coverage of the interferometer's
uv-plane.

All targets were observed in \hi\ line emission with a total
bandwidth of 16 MHz subdivided into 256 channels of 62.5 kHz
and, at the same time, in 1.344 GHz continuum with a bandwidth
of 128 MHz.

The individual observing runs were normally 11--13 hours long
(including time for calibration), leading to nearly complete
12 hour aperture syntheses. The on-source integration times
for each configuration and the totals are listed in
Table~\ref{atcaobs}.

\setcounter{table}{3}
\begin{table}
\caption{ATCA observation details. The array is operated in configurations with maximum 
baselines ranging from 75 m to 6 km. For 750-m, 1.5-km and 6-km configurations, 
4 subsets (A,B,C,D) each exist.}
\label{atcaobs}
\begin{tabular}{cllll}
\hline
Group           & Array          & Date       & Time        & Total \\
                &                &            & hh:mm       &  hh:mm       \\ \hline
SCG\,2159-3210  & 750D           & 1995-09-26 &   10:50     &   10:50 \\
SCG\,2353-6101  & 750D           & 1995-09-23 &    9:52     &    9:52 \\
SCG\,0018-4854  & 750D           & 1995-06-13 &    4:23     &   17:59 \\
                & 1.5B           & 1995-05-24 &    3:41     &         \\
                & 1.5D           & 1995-10-30 &   10:55     &         \\
SCG\,0122-3819  & 750D           & 1995-09-22 &   10:19     &   20:52 \\
                & 1.5C           & 2005-11-10 &   10:33     &         \\
SCG\,0141-3429  & 750D           & 1995-09-25 &    9:58     &    9:58 \\
SCG\,0227-4212  & 375            & 1995-08-15 &   10:27     &   10:27 \\ \hline
%GroupXX         & 750D & 1995-09-24 &   10:00 &   10:00 \\
\end{tabular}
\end{table}

Simultaneously, the 1.344 GHz continuum of the target galaxies
was also imaged in the second IF (Intermediate Frequency).
Because of the limited angular resolution of the resulting
images, and since only integral properties were used, these
images are not displayed here.

1934-638 was used as the primary flux and bandpass calibrator.
The adopted flux of 1934-638 is 14.94(15.01) Jy at 1.42(1.344)
GHz. The data reduction was performed in a standard fashion,
using the software package {\sc miriad} (cf. Miriad User's Guide)
and the underlying radio continuum was subtracted from the \hi\
line data in the uv-plane.

\hi\ data cubes were produced with channel widths of 125 kHz (i.e.
26.4 km s$^{-1}$). The \hi\ visibilities were weighted using a
robust factor of 0.5.

\section{Results}

In this section we present the results obtained from
our optical and radio pilot survey. We will first discuss 
the criteria used to confirm the group membership, the 
surrounding environment of the groups and their general
properties, like mass, crossing time, and radius.
Following this, we will discuss the properties of the
warm medium and we will present a description of the individual targets.
Within the individual targets, the properties of the member
galaxies will also be discussed.

\subsection{Group membership and environment}
All groups in our sample have been already confirmed spectroscopically,
assuming as {\it bona fide} criterion that all galaxies whose
velocity difference from the median systemic velocity of the groups
was less than $\pm$ 1000 \kms\ were group members. These
galaxies are called {\it concordant} galaxies and the minimum
number to confirm a group is 3 concordant galaxies.
In our sample only SCG2353-6101 has three confirmed galaxies, all other
targets have at least four or more concordant members.\newline

The groups examined here have been selected, among other criteria,
using an isolation criterion in the neighbourhood of the group
(for more details, see Hickson 1989 and I02). We searched for
neighbouring large scale structures using a 82$\arcmin$ radius,
which is equal to one Abell radius at the distance of the most distant group, SCG0122-3819.
In addition to this, we also add a redshift criterion: we assume
that a group is close to a cluster if the velocity difference
between the two is less than 3000 \kms, i.e. one order
of magnitude higher than the typical velocity dispersion of compact
groups and 2.5 times higher than the largest velocity dispersion
measured for clusters (1200 \kms, Zabludoff et al., 1993).

In Table~\ref{envi} we show the results of our search: within our constraints, 
no nearby cluster has been found, but two groups are not real compact
groups. SCG0122-3819 was identified by Katgert et al. (1996) as Abell 2911.
Looking at their catalog however, one realizes that there are three clusters with the
same name: the first, with 7 member galaxies, coincides with our group, at a redshift
z = 0.02. The second, which is in the original Abell catalog, is at redshift z = 0.08,
while a third one is at redshift z = 0.131. Following the literature until
present, we find that within an 82$\arcmin$ radius, SCG0122-3819 has a total of 23 galaxies 
at the same redshift. Twelve of them satisfy also the requirement of having  
$\Delta$m $<$ 3 from the brightest group galaxy, while the other seven do not
meet this requirement.\newline 
SCG2159-3210, also known as HCG90, has been studied in detail by Ribeiro et al. (1998), and defined 
as a {\it core+halo}
system, i.e. a loose group of galaxies with a central concentration. Two of the
galaxies in the loose group, {\it e} and {\it f}, satisfy the $\Delta$m $<$ 3 criterion, while 
the others, {\it g}, {\it i} and {\it o}, are fainter. \newline
The {\it core+halo} classification for SCG2159-3210
is consistent with the extended X-ray emission of the group, well beyond
the R$_{group}$ measured by Hickson. No X-ray data are available
at the moment for the other groups in our sample.\newline 
We must note however, that the search radius used by de Carvalho et al. (1997) 
and subsequently by Ribeiro et al. (1998) is significantly smaller than our own, 
15$\arcmin$ against 82$\arcmin$.
Restricting the search radius to 15$\arcmin$, to be able to compare our
findings with the cited work, we find that SCG0122-3819 has 7 additional galaxies,
5 of them fulfilling the $\Delta$m $<$ 3 criterion, while the other 2 are likely dwarf galaxies.\newline
All other targets are isolated systems, confirming the goodness of the original selection. 
We label as {\bf class A} all targets which can be considered really isolated on the
sky, with no obvious association to larger scale structure, while
we label as {\bf class B} the two groups which turned out to be
core+halo systems, SCG2159-3210 and SCG0122-3819.\newline
The number density of galaxies for all our groups, as estimated in Equation 1, 

\begin{equation}
\mathrm{\rho}  = \frac{3\mathrm{N}}{4\pi\mathrm{R^{3}}}
\end{equation}

is shown in column 4 of Table~\ref{envi};
in the specific cases of
SCG2159-3210 and SCG0122-3819, the density of galaxies has been estimated both for
the original galaxies in the SCGs catalog (4 for HCG90 and 5 for SCG0122-3819) and for 
all confirmed members within the 
15$\arcmin$ radius, see columns 4 and 5 of Table~\ref{envi}.

\setcounter{table}{4}
\begin{table*}
\begin{flushleft}
\caption{Classification of the compact groups and 
neighbouring large scale structures. The radius in Mpc
is the median radius of each group, while the value
given in parenthesis is the physical radius in Mpc
corresponding to 15$\arcmin$. Column 4 shows
the logarithm of the galaxy number density for the original galaxies in the SCGs, while in column
5 we show the logarithm of the galaxy number density obtained by including all galaxies within a
circle of 15$\arcmin$ radius. It is worth to note the remarkable galaxy density
of SCG0018-4854.}
\label{envi}
\begin{minipage}{\textwidth}
\begin{tabular}{lccccc}
\hline\hline
Group name   & Class & R              & Galaxy number density    & Galaxy number density (all galaxies) & Notes  \\ 
             &       & (Mpc)          & (gal $\times$ Mpc$^{-3}$) &  (gal $\times$ Mpc$^{-3}$)           &        \\ \hline
SCG2159-3210 & B     & 0.0436 (0.081) & 4.06                   &   3.60                                 & core+halo system \\
SCG2353-6101 & A     & 0.0841         & 3.08                   &   ---                                  &         \\
SCG0018-4854 & A     & 0.0232         & 4.88                   &   ---                                  &          \\
SCG0122-3819 & B     & 0.0708 (0.126) & 3.43                   &   2.77                                 & close to Abell 2911\footnote{Katgert (1996) 
identifies multiple redshift systems in this cluster, which at z = 0.08080}\\
SCG0141-3429 & A     & 0.1438         & 2.60                   &   ---                                  & HIPASS J0133-34 \\
SCG0227-4312 & A     & 0.2417         & 2.00                   &   ---                                  &  \\ \hline
\end{tabular}
\end{minipage}
\end{flushleft}
\end{table*}

\subsection{Internal dynamics and mass estimates}
Having confirmed as true compact groups all our targets, with the
exception of SCG2159-3210 and SCG0122-3819, we proceed to measure the characteristic
properties, i.e. three-dimensional velocity dispersion, crossing
time, mass, luminosity and mass-to-light ratio.

For velocity dispersion and crossing time we use the same
equations used in Hickson et al. (1992) and the results are
shown in columns 3 and 5 of Table~\ref{dyna}. 
$H_\mathrm{o}t_\mathrm{c}$, the dimensionless crossing time, spans from 0.017 to 0.058, with a median value of
0.024, which is larger than the value measured for HCGs, 0.016.

The median observed velocity dispersion is 178 \kms, while
the three-dimensional velocity dispersion is 315 \kms, in very good agreement
with what has been found for other compact group surveys, HCGs, and the
DPOSS II (Pompei et al., 2006).

For the mass estimate
we use different estimators, the virial and the projected mass. 
The expression for the virial mass is given in Equation 2, which
is valid only under the assumption of spherical symmetry.

\begin{equation}
\mathrm{M_{V}}  = \frac{3\pi\mathrm{N}}{2\mathrm{G}}
                  \frac{\mathrm{\Sigma_{i}V_{zi}^2}}{\mathrm{\Sigma_{i<j}1/R_{ij}}}
\end{equation}

where $R_{ij}$ is the projected separation between galaxies i and j,
here assumed to be the median length of the two-dimensional
galaxy-galaxy separation vector, corrected for cosmological effects. N is
the number of concordant galaxies in the system, and $V_{zi}^2$ the velocity component
along the line of sight of the galaxy {\it i} with respect to the centre
of mass of the group.
As observed by Heisler et al. (1985) and by Perea et al. (1990), the
use of the virial theorem produces the best mass estimates, provided
that there are no interlopers or projection effects.  In case one of
these two effects is present,  the current values can be considered an
upper limit to the real mass.

Another good mass estimate is given by the projected mass estimator, which is defined as:

\begin{equation}
\mathrm{M_{P}} = \frac{f_{P}}{GN} 
                 \mathrm{\Sigma_{i}V_{zi}^2}\mathrm{R_{i}}
\end{equation}

where $R_{i}$ is the projected separation from the centroid of the system,
and f$_{P}$ is a numerical factor depending on the distribution of the orbits
around the centre of mass of the system.

Assuming a spherically symmetric system for which the Jean's hydrostatic equilibrium
applies, we can express f$_{P}$ in  an explicit form (Perea et al., 1990). 
Since we lack information about the orbit eccentricities, we estimate the mass
for radial, circular and isotropic orbits and the corresponding expressions for
M$_{P}$ are given in Equations 4, 5 and 6 respectively:

\begin{equation}
\mathrm{M_{P}} = \frac{64}{\pi G}
                 {<V_{z}^2 R>}
\end{equation}

\begin{equation}
\mathrm{M_{P}} = \frac{64}{3\pi G}
                 {<V_{z}^2 R>}
\end{equation}

\begin{equation}
\mathrm{M_{P}} = \frac{64}{2\pi G}
                 {<V_{z}^2 R>}
\end{equation}
 
where R is the median length of the two-dimensional
galaxy-galaxy separation vector.\newline

The values for the virial mass and the projected mass estimator for
isotropic orbits are shown in cols. 6 and 7 of Table~\ref{dyna}: the different estimators agree
quite well with each other, with the exception of SCG0141-3429 and SCG0227-4312, where
the virial mass is higher than the other estimators. 
The discrepancy would point towards the possible presence of undetected interlopers or projection effects: to test
this, we re-measured all the mass estimators for these two groups, taking out one
galaxy each time. For SCG0141-3429, a very good agreement between the virial mass and the
other mass estimators is reached if galaxy {\it b} is discarded; the same is true for SCG0227-4312 by 
discarding galaxy {\it f}. While this might lead to think that these two galaxies are
likely interlopers, we must point out that the above estimators assume equal mass
among all group members. Using a luminosity weighted mass calculation for both groups,
and assuming log(M/L) = 0.82 (Roberts \& Haynes, 1994, readjusted for $\mathrm{h_0} = 0.67$),
we find that both virial and projected mass estimators agree very well within each other
for both groups: these values are the one quoted in parenthesis to the side of the respective
unweighted masses in Table~\ref{dyna}. The weighted values have been used for the estimate of the M/L ratio 
in column 9 of Table~\ref{dyna}.

The group masses vary from $1.02\times 10^{12}$ to $4.30\times 10^{13}$ $M_\mathrm{\sun}$, 
with a median value of $\sim$ $9.20\times 10^{12}$
$M_\mathrm{\sun}$, very similar to the values found for Hickson's compact groups.

\setcounter{table}{5}
\begin{table*}
\caption{Group dynamical properties. $\sigma_{r}$, R, $H_\mathrm{o}t_\mathrm{c}$  and M/L are
expressed in logarithmic values; mass and luminosity are given in solar units. 
All quantities for SCG0018-4854 and SCG0122-3819 have been estimated taking into
account 5 member galaxies rather than four, as in the original catalog; for a more detailed explanation, see Sect. 3.5.
The values in parenthesis for the  M$_{vir}$ and M$_{isotropic}$ are the luminosity weighted mass estimates.
The symbols follow Hickson et al. (1992)}
\label{dyna}
\begin{tabular}{l c c c c c c c c} \hline
Group name   &   Scale        & $\sigma_{r}$ & log(R) & $H_\mathrm{o}t_\mathrm{c}$ & M$_{vir}$         & M$_{isotropic}$      &  L & M/L   \\ 
             &(kpc/$\arcmin$) & (\kms)      & (kpc)  &                          & $M_\mathrm{\sun}$    & $M_\mathrm{\sun}$   & $L_\mathrm{\sun}$     &       \\ \hline
SCG2159-3210 & 11.34          & 1.997       & 1.640  & -1.649                   & $1.89\times 10^{12}$ &$1.02\times 10^{12}$ & $6.20\times 10^{10}$ & 1.22 \\
SCG2353-6101 & 19.50          & 2.379       & 1.925  & -1.758                   & $1.58\times 10^{13}$ &$1.14\times 10^{13}$ & $3.57\times 10^{10}$ & 2.50 \\
SCG0018-4854 & 14.52          & 2.065       & 1.614  & -1.735                   & $3.03\times 10^{12}$ &$1.31\times 10^{12}$ & $4.07\times 10^{10}$ & 1.51 \\
SCG0122-3819 & 16.67          & 2.392       & 2.292  & -1.405                   & $4.19\times 10^{13}$ &$1.81\times 10^{13}$ & $9.78\times 10^{10}$ & 2.27 \\
SCG0141-3429 & 25.92          & 2.119       & 1.994  & -1.418                   & $1.43\times 10^{13}$ ($6.96\times 10^{13}$)&$6.19\times 10^{12}$ ($7.53\times 10^{13}$)& $3.22\times 10^{10}$ & 2.37 \\
SCG0227-4312 & 22.81          & 2.452       & 2.383  & -1.374                   & $1.27\times 10^{14}$ ($3.98\times 10^{13}$)&$4.60\times 10^{13}$ ($4.30\times 10^{13}$)& $1.44\times 10^{11}$ & 2.47 \\ \hline
\end{tabular}
\end{table*}

Group luminositites were derived by summing up the luminosity of all member
galaxies. For consistency with the published works on HCGs, we estimated B-band luminosities:
our imaging data are in b$_{j}$ (from COSMOS scans) or in R (from the observations); we converted 
all magnitudes to standard
B using the equation (Prandoni et al., 1994):

\begin{equation}
B  =  b_{j}+0.23\times(B-V)
\end{equation}

and assuming standard (B-V) colors depending on the morphological
type of the member galaxies, going from (B-V)=0.9 for ellipticals
to (B-V)=0.5 for irregulars. 
Given the closeness of the targets, morphological {\it k} corrections were neglected;
the derived values for the luminosity are given in column 8 of Table~\ref{dyna}.

As a reference for the solar magnitude, we used the paper by Jorgensen
(1994). From stars with a $(B-V) \sim 0.65$ (i.e. the same colour as
the Sun) and assuming $M_\mathrm{R,\sun}$ = 4.42
(Binney \& Merrifield, 2001), we obtain $M_\mathrm{B,\sun}$ = 5.48.

From the median group mass and our estimated B band luminosity, we derive the mass-to-light ratio, which 
range from 16 to 316, with a median value of 188. These values are higher than for HCGs, but comparable 
with the values found for loose groups.

It is interesting to note that for the only group we have in common with
Hickson's work (HCG90 = SCG2159-3210), we agree extremely well with his estimates 
for velocity dispersion, crossing time, mass and luminosity.

\subsection{Properties of the warm medium}
From the \hi\ observations described in Sect. 2.2 we derive the following
pieces of information on individual galaxies within each of the
six groups tabulated in Table 1:
\begin{itemize}
\item{} \hi\ emission distributions;
\item{} \hi\ systemic velocities,
\item{} \hi\ line widths,
\item{} \hi\ total line flux densities,
\item{} \hi\ velocity fields,
\item{} \hi\ continuum emission
\end{itemize}
From these data we are able to derive the \hi\ mass,
given in Equation 8,

\begin{equation}
M(HI) = 2.356 \times 10^5\ D^2 \times f(HI)\ [$\msol$]\quad,
\end{equation}

where $D$ is the distance in Mpc.\newline We next derive the HI gas
deficiency, by comparing with the work of Haynes \& Giovanelli
(1984) on isolated galaxies.

We define the \hi\ deficiency as:

\begin{equation}
Def(HI) = log[M(HI)_{pred}]-log[M(HI)_{obs}]
\end{equation}

where M(HI)$_{pred}$ is the predicted total \hi\ mass for the group,
obtained by summing up all the \hi\ mass of the member galaxies, and assuming that these 
are field galaxies of the same Hubble type observed in the group.
M(HI)$_{obs}$ is the observed total \hi\ mass of the group, obtained by summing up all the
\hi\ mass measured in the member galaxies.

Values pertaining to the determination of \hi\ deficiencies are collected in Table~\ref{hi_def},
while general \hi\ properties of the member galaxies are tabulated in Table~\ref{scg_hi}.

We stress that the \hi\ deficiency calculated here makes use
of the \hi\ mass, corrected by {\it h$^{2}$}, without any normalization
for luminosity or diameter. In addition, we assumed that elliptical
galaxies, such as SCG2159-3210B and SCG2159-3210C, have no \hi\ gas, while
we assigned to S0 galaxies their face value, as given in Table IV
of Haynes \& Giovanelli (1984), without making any assumption on their possible
past as spiral galaxies.
The standard estimate of the error, or {\it s.e.e.}, has been calculated
by averaging together the individual {\it s.e.e.} for the member galaxies of each
group.

Many compact groups (see VM01, for
an example) are considered \hi\ deficient, i.e. they
contain less neutral gas than one would expect to find
when summing up the average \hi\ mass of field galaxies of the
same Hubble type as the group members.

We find that groups in our sample have an average \hi\ deficiency of
$Def(HI) = 0.80$, see Table~\ref{hi_def}; if we exclude SCG2159-3210, which is a special case,
given the presence of two elliptical galaxies and a Sy2 nucleus in one of
its members,
we find that the average \hi\ deficiency is 0.60, in
relatively good agreement with the results by VM01.

\setcounter{table}{6}
\begin{table}
\caption{Measured values of \hi\ deficiency for SCGs in our sample. Following VM01, we assume that a group
can be considered \hi\ deficient if the difference between predicted and
observed mass is more than twice the standard estimate of the error.}
\label{hi_def}
\begin{tabular}{l c c c c c} \hline
Group name   &   M(HI)$_{pred}$     & M(HI)$_{obs}$       & s.e.e. & Def$_{HI}$ \\ 
             & $M_\mathrm{\sun}$    & $M_\mathrm{\sun}$    &       &            \\ \hline
SCG2159-3210 & $10.44\times 10^{9}$ & $0.29\times 10^{9}$  & 0.30  & 1.57       \\
SCG2353-6101 & $19.46\times 10^{9}$ & $3.77\times 10^{9}$  & 0.24  & 0.71       \\
SCG0018-4854 & $23.15\times 10^{9}$ & $7.21\times 10^{9}$  & 0.24  & 0.51       \\
SCG0122-3819 & $24.67\times 10^{9}$ & $15.94\times 10^{9}$ & 0.29  & 0.19       \\
SCG0141-3429 & $26.75\times 10^{9}$ & $4.10\times 10^{9}$  & 0.24  & 0.81       \\
SCG0227-4312 & $44.16\times 10^{9}$ & $6.76\times 10^{9}$  & 0.26  & 0.82        \\ \hline
\end{tabular}
\end{table}

For galaxies with SF-related activity (after exclusion of AGNs,
LINERs or other types of Emission Line Galaxies [ELGs]) one
can derive supernova rates, $\nu_{\rm SN}$ from the measured
radio continuum total flux densities according to

%
% flux to radio power conversion:
%
% 1 Jy = 10^-23 erg s^-1 cm^-2 Hz^-1 = 10^-26 W m^-2 Hz^-1
%
% P [W Hz^-1] = 4 pi D^2 [m^2] x S [W m^-2 Hz^-1]
%             = 1.1965e20 x D [Mpc] x S [Jy]
%

\begin{equation}
P_{\rm tot}\ [{\rm W\,Hz}^{-1}]\ =\ f_{\rm SN} \times \nu_{\rm SN}\
[{\rm yr}^{-1}]\quad.
\end{equation}

In Eq. 10 $P_{\rm tot} = 4 \pi\,D^2\,f(1.34)$ is the measured total radio
power derived from the radio continuum flux density, $f(1.34)$, and
$f_{\rm SN}$ is a proportionality factor:

\begin{equation}
f_{\rm SN}\ =\ ({\nu \over 1.5\,GHz})^{-{x \over 2}}\
({B \over 5\,\mu G})^{{x-2 \over 2}}\ \times 6.4\times10^{22}
\quad,
\end{equation}

in which {\it x} is the radio continuum spectral index. 

For a fiducial magnetic field strength of $B = 5 \mu$G and a spectral
index of x = 2.2, the resulting factor is
$f_{\rm SN} = 7.2(2173) \times 10^{22}$ at the observed frequency of
1.344 GHz.
$P_{\rm 21}^{1.34} = 10^{21}\ P_{\rm tot}$ is the total radio power
at 1.34 GHz in units of $10^{21}$ W Hz$^{-1}$. This value was used
for all galaxies listed in Table~\ref{scg_rc}.

Based on the $\nu_{\rm SN}$ values one can calculate star formation
rates ({\it SFR}s) for a given initial mass function (IMF). For a
Salpeter IMF with an exponent of --2.35, the following relationship
exists:

\begin{equation}
SFR\ =\ \nu_{\rm SN}\ \times\ {2.35 \over 1.35}\ \times\
{{(M_{\rm up}^{-0.35}-M_{\rm low}^{-0.35})} \over
{(M_{\rm up}^{-1.35}-M_{\rm SN}^{-1.35})}}
\quad.
\end{equation}

For an upper and lower limit for the stellar mass range of
$M_{\rm low}$ = 1 \msol\ and $M_{\rm up}$ = 100 \msol\ and a
minimal mass at which a supernova can explode of $M_{\rm SN}
= 8$ \msol, this leads to the simple relation

\begin{equation}
SFR\ =\ 23.9 \times\ \nu_{\rm SN} $\msolyr$ \quad.
\end{equation}

Assuming a value of SFR = 0.5\msolyr for quiescent galaxies
(Kennicutt et al., 2005), and a threshold for a starburst: SFR $>$ 2.0\msolyr, 
we find that 6 galaxies out of 28 in our sample are star forming galaxies and 3 of these
have ongoing starburst.

\subsection{Individual targets}
\begin{itemize}
\item{}{\bf SCG2159-3210 (HCG90)}: this is a group with four confirmed member galaxies
with measured optical recession velocities, named here {\it a}
through {\it d}. Morphologies for these galaxies were obtained 
from Mendes de Oliveira et al. (1994) and from Hickson et al. (1989), 
as we lack an optical image for this group.
Five additional galaxies were observed by De Carvalho et al. (1997), for a total
of 9 confirmed members; however in this paper we consider
only the original 4 galaxies included in Hickson's catalog and in our own.\newline
This is a group with mixed morphology, with one spiral galaxy, {\it a} (=NGC 7172), which is a Sa,
two ellipticals, {\it b} and {\it c} and one irregular, {\it d}. 
Galaxies {\it a} through {\it d} have
X-ray emission (Mulchaey et al., 1998); galaxy {\it a} 
is a Sy2 (Coziol et al., 1998), while galaxies {\it b} and {\it c}
have no optical line emission at all and galaxy {\it d} is a LINER.

An additional diffuse X-ray component, with T $\sim$ 0.7 KeV, has been observed
(Mulchaey et al., 1998; White et al., 2003), but its distribution
at the moment is not very clear. In Mulchaey et al., the X-ray 
isophotes, which reach down to 2$\sigma$ above the background, extend
along the N-W axis of the group, almost reaching galaxy {\it a}, see his Fig. 1k.
In the data shown by White the X-ray emission is centered on galaxies {\it b}, {\it c}
and {\it d}, with an extension towards W.
 
Two of the four original members of the group, {\it b} (=NGC 7176)
and {\it d } (=NGC 7174), are strongly interacting with each other;
galaxy {\it d} shows two tidal plumes, while  galaxy {\it b} shows a 
luminous bridge towards {\it d} and distorted isophotes (Longo
et al., 1994). White et al. (2003) demonstrate that a significant
amount of optical light, $\sim$ 45$\%$ of the total, is in the form
of a diffuse component distributed between galaxies {\it b}, {\it c}
and {\it d}.

Galaxies {\it a} and {\it d} have been observed in CO (Boselli et al., 1996) and the estimated mass of 
H$_{2}$ is of the order of 10$^{8.8} M_{\odot}$.
The ratio between the M(H$_{2}$) and the galaxy blue luminosity is an indication
of how much a galaxy has been disturbed by some kind of interaction (see again Boselli et al., 1996): 
we measure log(M$_{H_2}$/L$_{B}$)=-1.24 for galaxy {\it a},
and log(M$_{H_2}$/L$_{B}$)=-0.91 for galaxy {\it d}. The first value is consistent
with an unperturbed galaxy, while the second is more typical of disturbed galaxies.
This correlates very nicely with the morphology shown in the optical image, where galaxy {\it d} is strongly
interacting with {\it b}.

Only galaxy {\it a} is detected in \hi\ line emission, see  
Fig.~\ref{fig:scg3hi}: the gas distribution is highly asymmetric, with  more gas in
the eastern half of the galaxy disk than in the west.

\setcounter{figure}{1}
\begin{figure*}
\includegraphics[scale=0.8]{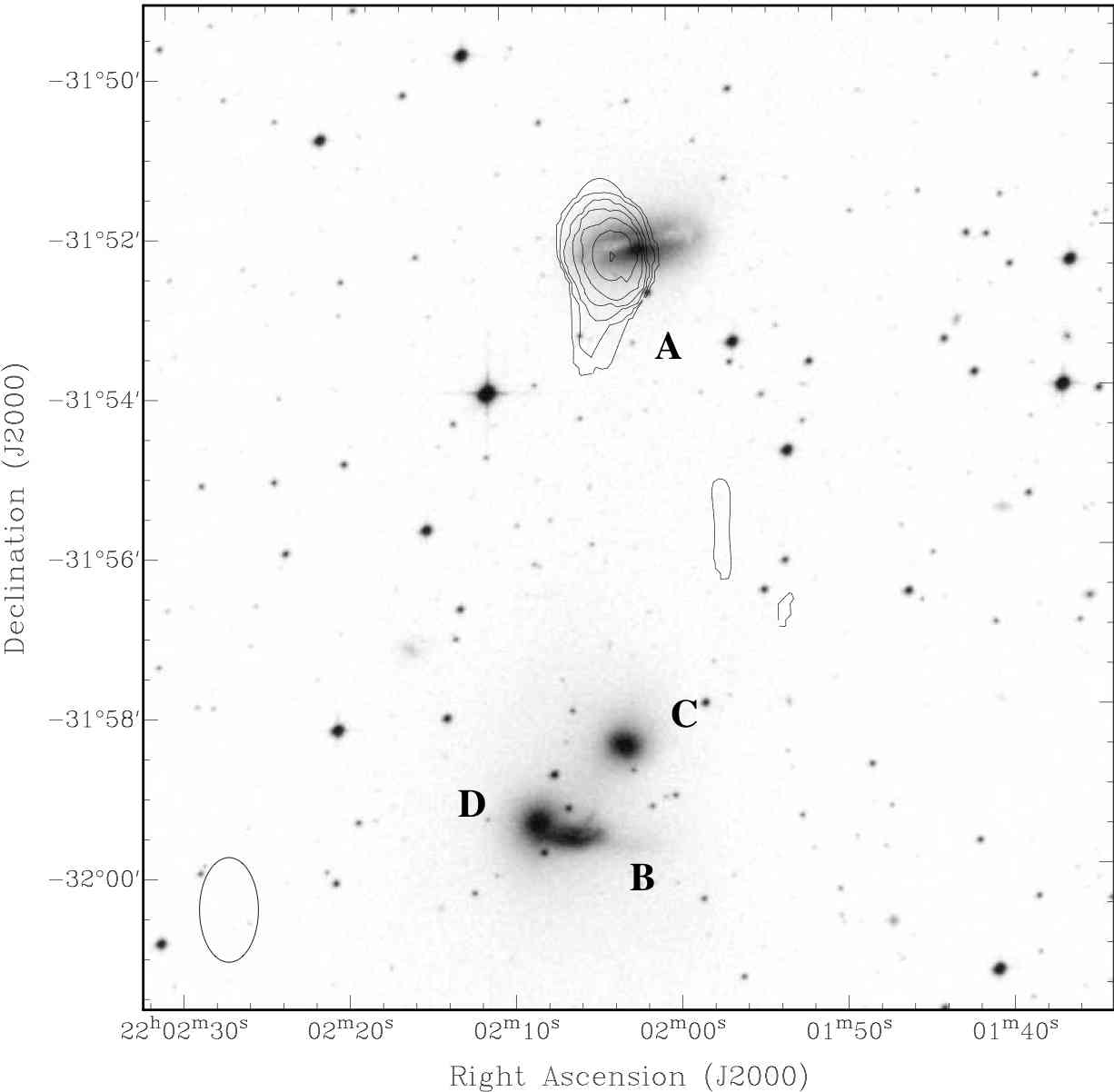}
\caption{ATCA \hi\ image of SCG\,2159-3210, superimposed on a red DSS-2
optical plate. The contour levels are at 0.12, 0.17, 0.24, 0.34, 0.48
0.68, and 0.96 Jy km s$^{-1}$ beam$^{-1}$. The field of view is
$12\farcm2\times 12\farcm6$ and the angular resolution ({\it
FWHM}=$80\arcsec \times 45\arcsec$) is indicated in the lower left. N is up
and E to the left}
\label{fig:scg3hi}
\end{figure*}

Due to the presence of strong radio continuum emission from the
nuclear area and beam smearing effects, no significant \hi\
emission is seen in the western disk. Higher angular resolution
is required to reduce this effect.
Clearly the \hi\ line flux, {\it f(HI)}, is reduced by the
absorption feature, which leads to an underestimate of the total
\hi\ gas mass, {\it M(HI)}.

The \hi\ emission distribution in the eastern half is disturbed,
with a tidal tail protruding towards south, i.e. in the direction
of the other group members.
All other galaxies show no \hi\ line emission at the sensitivity
limit of the current data, but all of them show continuum emission:
it was not possible to determine the star formation rate (SFR) from radio continuum data
for galaxy {\it a} because the contribution of its Sy-2 AGN cannot
be assessed. Galaxy {\it d} however is the strongest continuum emitter, with an estimated
star formation rate of  0.92 \msolyr\, see Table~\ref{scg_rc}.
It is likely that the gas reservoir which maintains the star formation comes from
the molecular gas, plus cold \hi\ gas, as picked up in absorption against the
nuclear emission. An approximate estimate
of the gas mass is 10$^{9} M_{\odot}$, which should be enough for
the central star formation activity for a limited amount of time.

The dispersion of the optical recession velocities of the four
group members is only ~ $\sim$ 100 \kms, which can make tidal interactions
very efficient at removing gas from galaxies.

The two-dimensional velocity field of galaxy {\it a} (not
displayed) shows little substructure, because most of the gas
detected is located on one side of the differentially rotating
part of the galaxy disk, with almost constant rotational velocities.

\item{}{\bf SCG2353-6101}: this is a small group with only 3 concordant member
galaxies; galaxies {\it a} and {\it b} are very close to each other, and likely
interacting. The optical halo of the two galaxies
is touching in the middle, so masking of one of the galaxies is needed
for a good photometric decomposition of the other.\newline
Galaxy {\it a} displays two extended spiral arms, which close on each other at
a radius of $\sim$ 46$\arcsec$; a strong isophotal twist is observed.
Galaxy {\it b} is a diffuse spiral, with HII regions and an almost invisible
bulge, while galaxy {\it c} shows a boxy bulge and an exponential disk.
A good photometric decomposition is obtained for all three targets
using no more than two or three components, see Table~\ref{scg_pho}.
%Galassia A: Sersic con expdisk+ estese braccia a spirale, con twist.
%Galassia B: almost pure exponential, with substructures, point-like.
%Galassia C: Sersic con expdisk.

A literature search shows that both galaxies {\it a} and {\it b} have been
observed in X-ray, but the data enable only an upper limit
determination, with L$_{X} <$ 10$^{40.93}$ erg s$^{-1}$ and L$_{X} <$ 10$^{41}$ erg s$^{-1}$
for galaxy {\it a} and {\it b} respectively (see Burstein et al., 1997).

Galaxy {\it a} is a star-forming galaxy, with an optical SFR of 1.2 M$\odot$ yr$^{-1}$
(Crocker et al., 1996), which is also confirmed by Coziol et al., 2000.
Its H$\alpha$ luminosity is L$_{H\alpha+[NII]}$ = 10$^{40.83}$ erg s$^{-1}$,
comparable to the X-ray upper limit. However the galaxy is only marginally
detected in \hi, as can be seen in Fig.~\ref{fig:scg6hi}.

\onlfig{3}{
\begin{figure*}
%\hspace*{-20mm}
\includegraphics[scale=0.8]{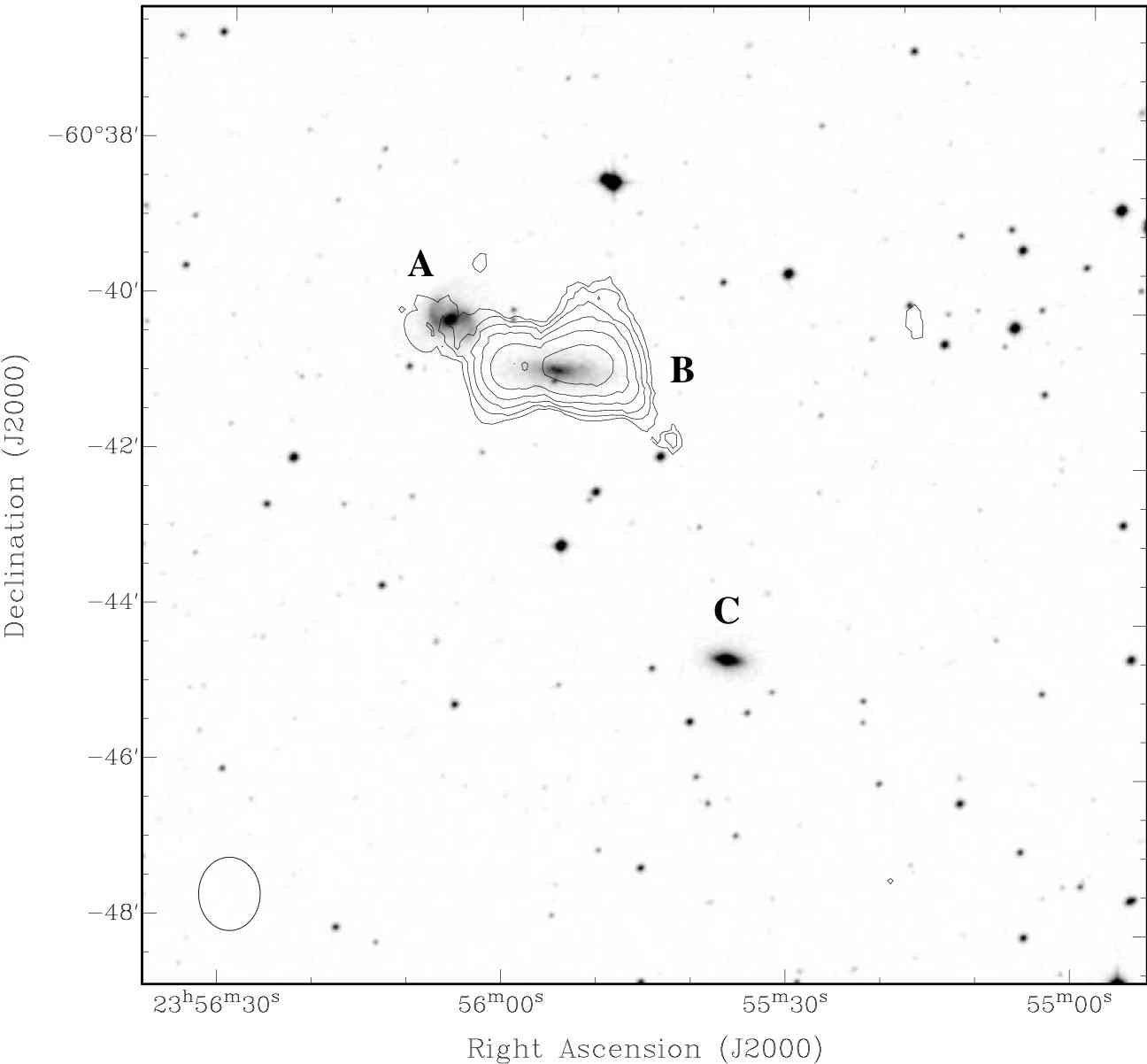}
\caption{ATCA \hi\ image of SCG\,2353-6101, superimposed on a red DSS-2
optical plate. The contour levels are at 0.24, 0.34, 0.48, 0.68, 0.96
and 1.36 Jy km s$^{-1}$ beam$^{-1}$. The field of view is
$13\farcm0\times 12\farcm6$ and the angular resolution ({\it
FWHM}=$50\arcsec \times 42\arcsec$) is indicated in the lower left.
N is up and E to the left}
\label{fig:scg6hi}
\end{figure*}
\afterpage{\clearpage}
}

Galaxy {\it b}, on the other hand,  has a lot more \hi\ gas; a tidal tail is found to
emanate from the western end of its disk towards
north. 

The extended \hi\ distribution in galaxy {\it b} and the almost overlapping optical halo
between galaxies {\it a} and {\it b} might hint towards
a past interaction between the two galaxies, which would have swept most of the
gas away from the main galaxy and heated up the small remaining quantity,
which is now emitting X-rays. On the other hand, the difference in recession velocity of 
the two galaxies for which \hi\ measurements are available, $\sim$ 600 \kms, is high (see
Fig.~\ref{fig:scgvel_2353}). 

\onlfig{4}{
\begin{figure*}
%\hspace*{-20mm}
\resizebox{\hsize}{!}{\includegraphics{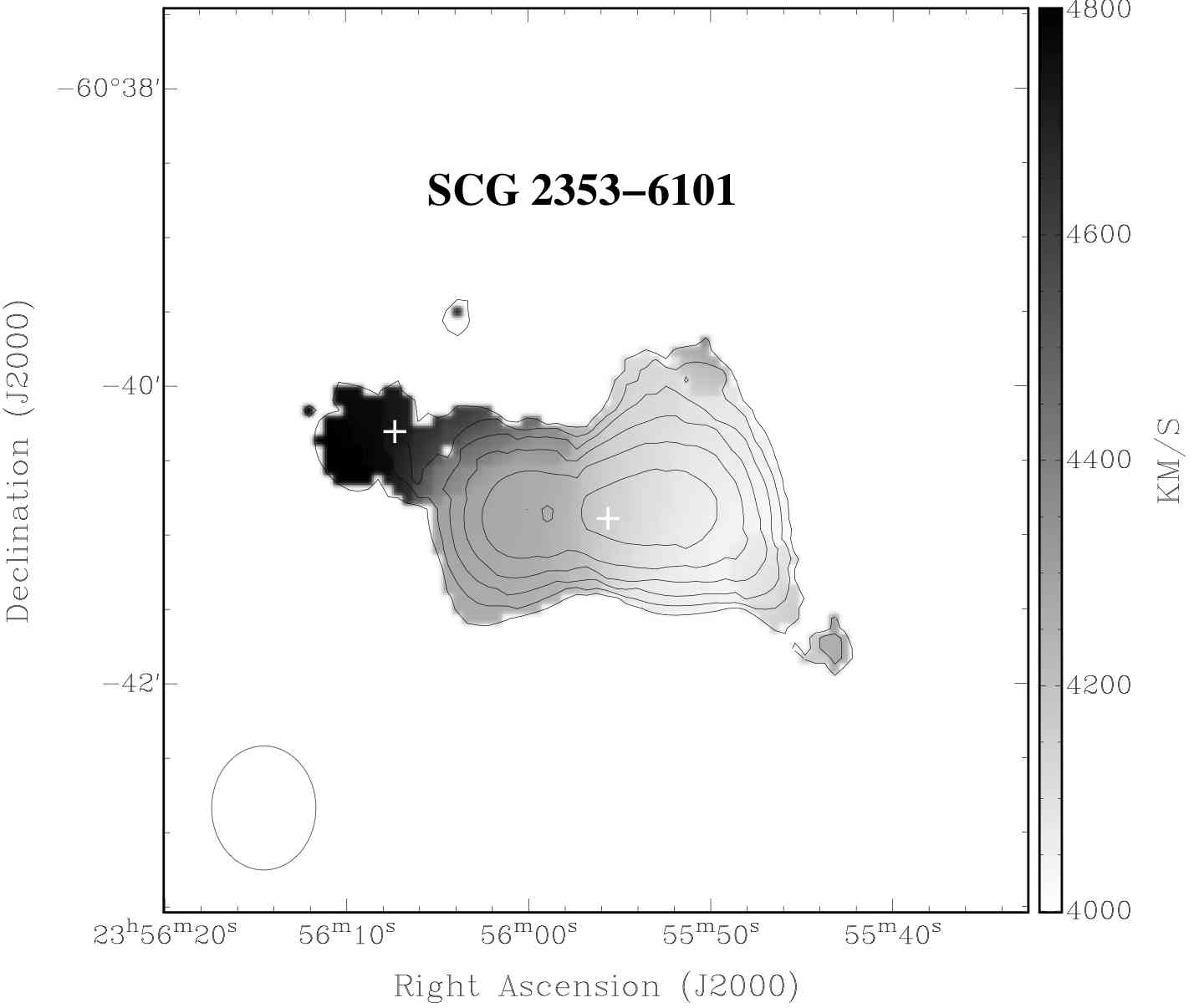}}
\caption{ATCA \hi\ image of SCG\,2353-6101 superimposed on the
\hi\ velocity field, the white crosses
indicate the center of galaxy {\it a} and {\it b}. The contours are the same as in
Fig.~\ref{fig:scg6hi}, the field of view is $5\farcm9\times
6\arcmin1$ and the angular resolution ({\it FWHM}= 50$\arcsec
\times$42$\arcsec$) is indicated in the lower left. The velocity
scale is shown on the right and the orientation is N up and E to
the left.}
\label{fig:scgvel_2353}
\end{figure*}
}

It is not clear whether we can assume the existence of a physical
gaseous bridge between {\it a} and {\it b} ; what
looks like a steady velocity gradient in the data might be caused
by beam-smearing, rather than a gas flow. \newline
The last remaining galaxy in this group, which can be
seen in the south-western corner of Fig.~\ref{fig:scg6hi}, does not
exhibit any \hi\ emission down to the sensitivity limit of the
current data, see Table~\ref{scg_hi}.

No good-quality radio continuum data are available due to severe
confusion of the field-of-view by a nearby radio galaxy with
a total flux density of $>5$ Jy. Visual inspection of the
resulting image shows radio continuum emission from the centre
of galaxy {\it b}, but the data quality is too poor to
further quantify this.

\item{}{\bf SCG0018-4854}: this group has four galaxies in close proximity, with a fifth
member at a larger projected distance, 15$\arcmin$.
All four close members show a disturbed morphology and a late
type appearance: the brightest galaxy, {\it a} (=NGC\,92), has an extended tidal tail, with many HII regions
very prominent in H$\alpha$, a bright circumnuclear ring of
star formation, with a pronounced knot and a strong central starburst, with a SFR, as determined from
the ATCA radio continuum observations, of 14.7 \msolyr\ and a SN
rate of about 0.6 yr$^{-1}$. The galaxy has a strong
central radio continuum source, which causes the absorption hole
visible in Fig.~\ref{fig:scg9hi}. 

\onlfig{5}{
\begin{figure*}
%\hspace*{-20mm}
\includegraphics[scale=0.8]{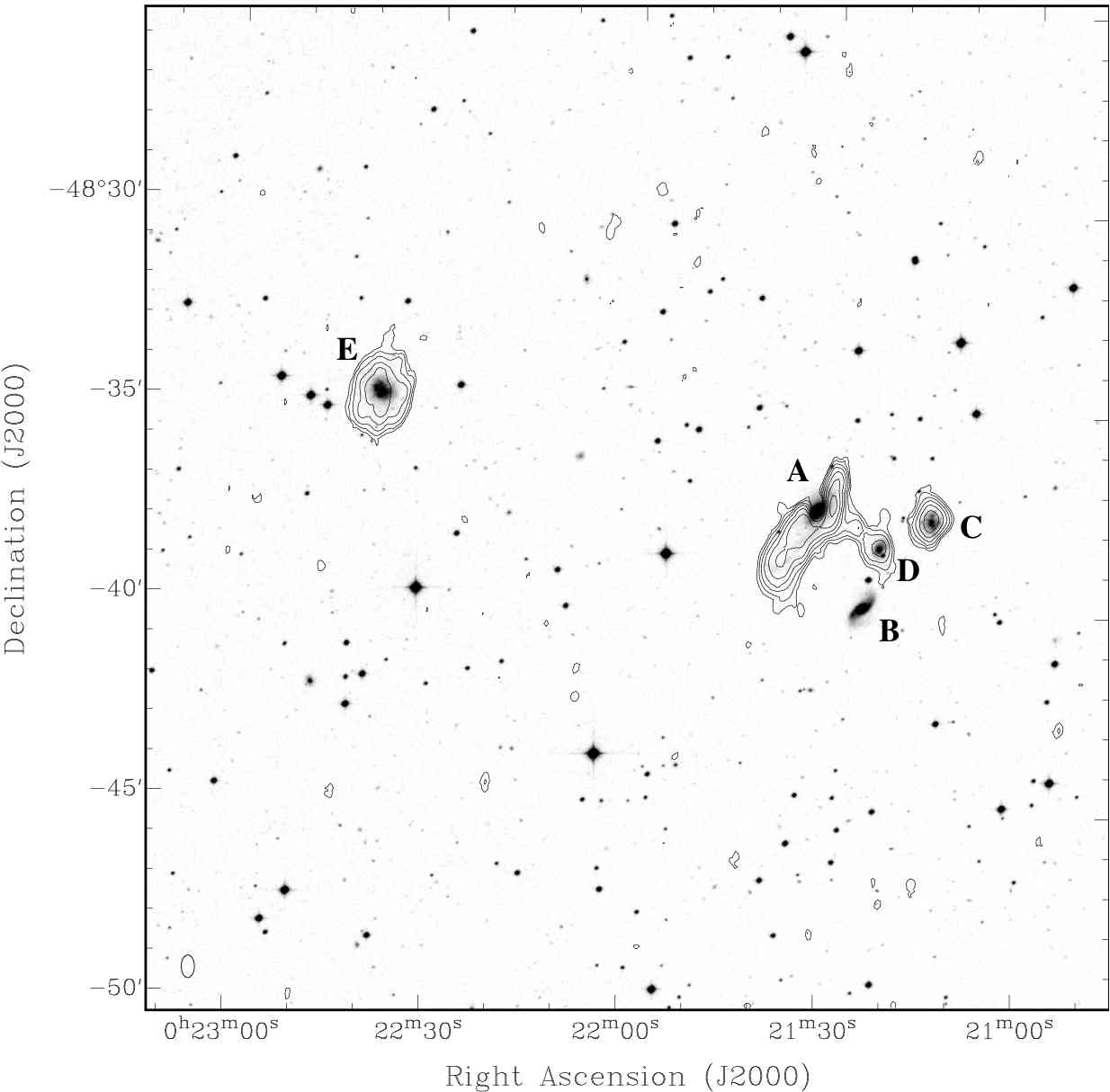}
\caption{ATCA \hi\ image of SCG\,0018-4854, superimposed on a red DSS-2
optical plate. The contour levels are at 0.12, 0.17, 0.24, 0.34, 0.48,
0.68 and 0.96 Jy km s$^{-1}$ beam$^{-1}$. The field of view is
$24\farcm2\times 25\farcm2$ and the angular resolution ({\it
FWHM}=$38\arcsec \times 22\arcsec$) is indicated in the lower left.
N is up and E to the left}
\label{fig:scg9hi}
\end{figure*}
}

\hi\ gas is clearly detected in its disk and
along the whole length of its southern tidal tail.

Galaxy {\it b} (=NGC\,89) has strongly twisted isophotes, a
Sy2 nucleus and a NE oriented jet, well visible on H$\alpha$ images (G. Temporin, private
communication). It is the only galaxy in the group
not being detected in \hi\,  with an upper limit of
$3\times10^7$ \msol\ . However we detected radio continuum emission  
(Table~\ref{scg_rc}), albeit fainter than galaxy {\it a}.

Galaxy {\it c} (=NGC\,87) is an irregular galaxy, with no real centre and strong
star formation.

Galaxy {\it d} (=NGC\,88) has another active nucleus (Coziol et al., 2000);
its \hi\ emission forms a bridge towards galaxy {\it a}.

The fifth galaxy, {\it e} (=ESO198-G013), is a spiral galaxy, with diffuse arm morphology;
\hi\ line emission is detected, but the gas distribution does not show any major
disturbances.

Surface photometry of the galaxies has been possible only with extensive
masking of nearby objects and irregular features, the results are listed
in Table~\ref{scg_pho}, while \hi\ data are given in
Table~\ref{scg_hi}.

The relative projected velocities of the member galaxies of this group 
are small, resulting in a group velocity dispersion of only 120 \kms.
A large fraction of this value is contributed by galaxy {\it e}, which
has a relative projected velocity that differs from the rest by about 200 \kms. At
the same time it is also at a large projected distance from the
core of the group.

Fig.~\ref{fig:scgvel_0018} shows that the rotational amplitude of
galaxy {\it a} encompasses the total velocity range
observed in both companions detected in \hi\ line emission,
{\it c} and {\it d}, and
the gas bridge connecting them. The NW half of galaxy {\it a} is the
receding side. The velocities of the \hi\ gas in
the tidal tail are an extension of those observed on that side
of the galaxy disk, clearly suggesting a physical link between
disk gas and tail.

\onlfig{6}{
\begin{figure*}
%\hspace*{-20mm}
\resizebox{\hsize}{!}{\includegraphics{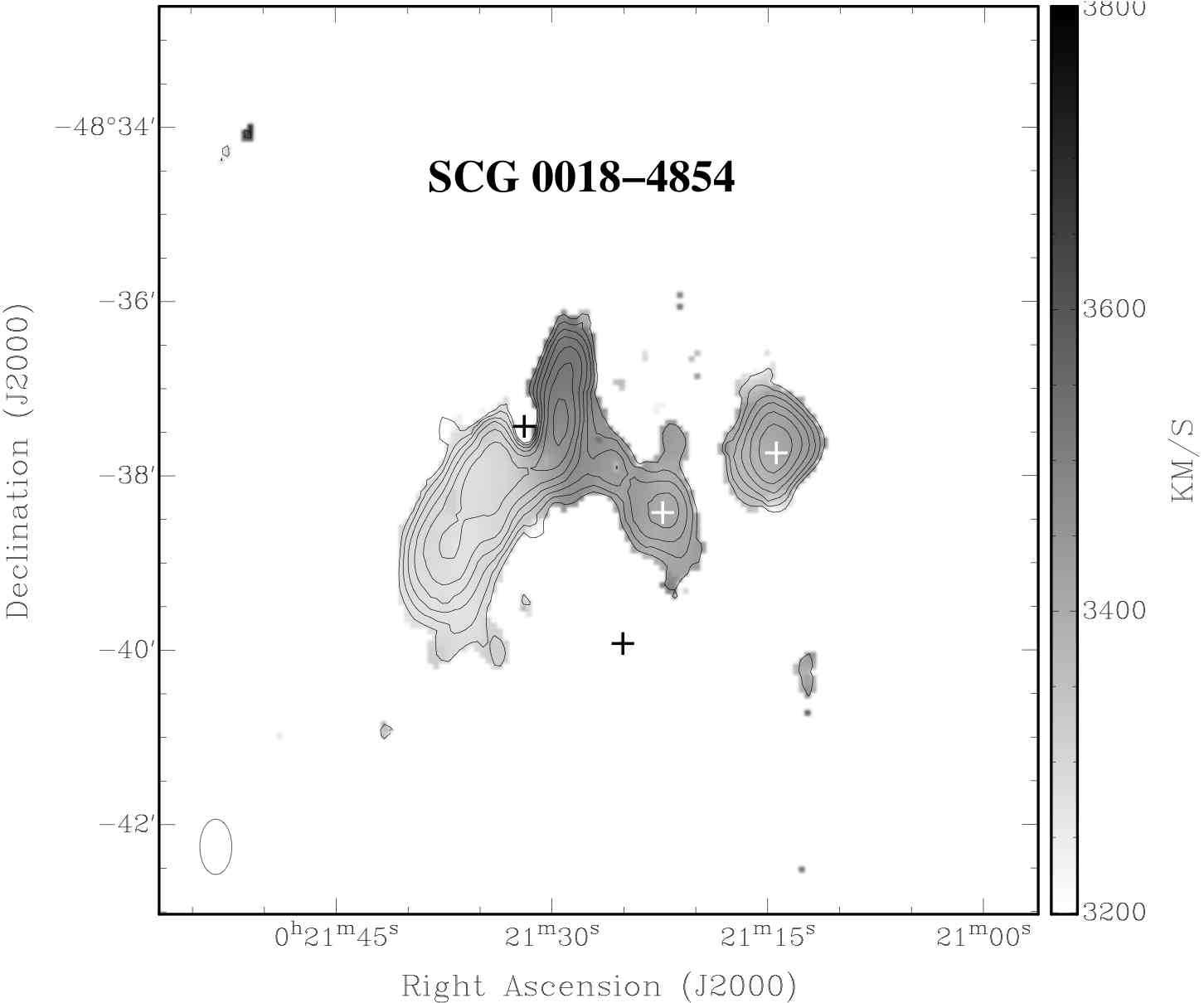}}
\caption{ATCA \hi\ image of SCG\,0018-4854, superimposed on the
\hi\ velocity field. The contours are the same as in
Fig.~\ref{fig:scg9hi}, the field of view is $10\farcm2\times
10\arcmin5$ and the angular resolution ({\it FWHM}= 38$\arcsec
\times$22$\arcsec$) is indicated in the lower left. Crosses indicate
the center of each member galaxy, the velocity
scale is shown on the right, and the orientation is N up and E to the
left.}
\label{fig:scgvel_0018}
\end{figure*}
}

X-ray observations of the close members show X-ray emission
mainly centered on galaxies {\it a} and {\it b} (G. Trinchieri, private communication),
with a possible bridge between the two galaxies, unfortunately at a level
comparable to the background. It is interesting to note an
anti-correlation between \hi\ and X-ray emission, with two separate
bridges of hot and warm gas connecting galaxy {\it a} with {\it b} and {\it d} respectively.\newline
The \hi\ velocity field of galaxy {\it e} (=ESO\,194-G\,013; not
displayed) is regular. The north-eastern side of the galaxy is
the receding part.

All galaxies have been observed in CO by Boselli et al. (1996): galaxies {\it a}
and {\it b} contain significant amount of M(H$_{2}$), 10$^{9.3}$ M$_{\odot}$ and 10$^{8.6}$ M$_{\odot}$,
respectively; surprisingly enough galaxy {\it a} has not been detected at all at 60$\mu$m from IRAS.

The $log(M_{H_{2}}/L_{B})$ is equal to -0.068 and -0.84 respectively for galaxies {\it a} and {\it b}:
according to these measures, both galaxies are perturbed.
Galaxies {\it c} and {\it d} instead have negligible amounts of molecular gas.

With two active nuclei and a strong starburst, plus two star-forming galaxies,
this group is the likely southern counterpart of the famous HCG 16, the youngest
and most active group in Hickson's catalog, except at an higher density.
%
%galassia A: Sersic+Sersic+expdisk (DV+expdisk)
%galassia B: expdisk
%galassia C: Sersic+Sersic (expdisk)
%galassia D: Sersic+expdisk
%galassia E: un caos for nearby stars stripes and diffuse light: Sersic+expo!!

\item{}{\bf SCG0122-3819}: this group lies within 10$\arcmin$ radius from the cluster Abell 2911, 
which counts 31 galaxies in total (Katgert et al., 1996), but it is
at a completely different redshift. Galaxies
{\it a}, {\it b} and {\it d} are barred, while galaxy {\it c} shows
an extended tidal tail, very similar to that observed in SCG0018-4854{\it a}. 
Galaxies {\it a} and {\it c} are very close to
each other and likely interacting: an halo of optical light between
the two is clearly visible between the two objects. 

A boxy bulge for galaxy {\it d} was detected by L\"utticke et al.(2000),
but the detailed surface photometry executed with GALFIT shows that
this is really an exponential bar. Actually this group is dominated by
barred galaxies, {\it a} and {\it b}, showing a flat profile (i.e. strong)
bar, while {\it d} has an exponential one.

\onlfig{7}{
\begin{figure*}
%\hspace*{-20mm}
\resizebox{\hsize}{!}{\includegraphics{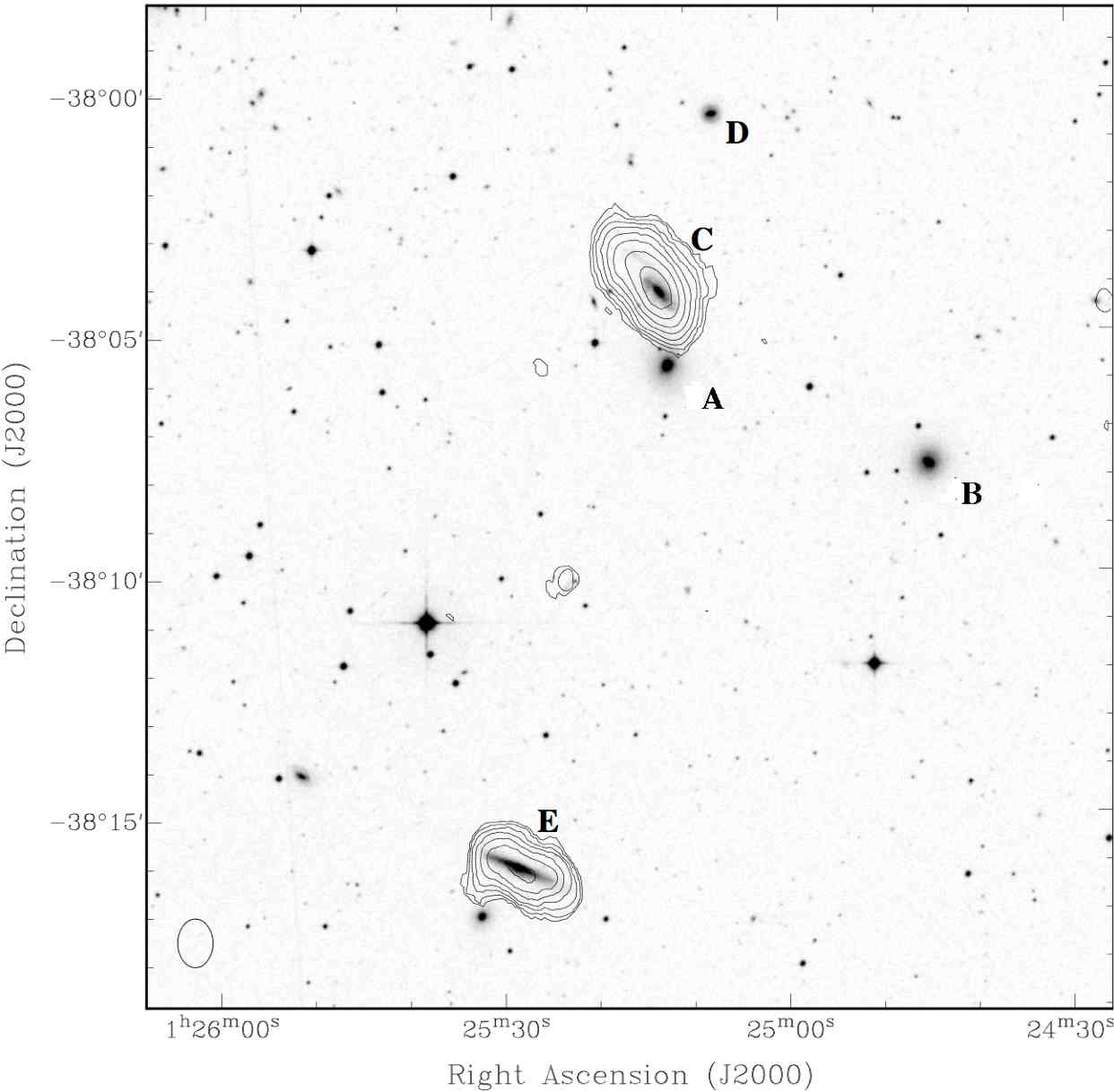}}
\caption{ATCA \hi\ image of SCG\,0122-3819, superimposed on a red DSS-2
optical plate. The contour levels are at 0.2 0.28 0.4 0.56 0.8 1.12 1.6
2.24 Jy km s$^{-1}$ beam$^{-1}$. The field of view is
$20\farcm2\times 20\farcm9$ and the angular resolution ({\it
FWHM}=$38\arcsec \times 22\arcsec$) is indicated in the lower left.
N is up and E to the left}
\label{fig:scg11hi}
\end{figure*}
\afterpage{\clearpage}
}

Of the four optically identified member galaxies in SCG\,0122-3819
only one, galaxy, {\it c}, is also detected in \hi\ line emission
(Fig.~\ref{fig:scg11hi}). This is a starburst, with a 
radio continuum SFR of approximately 2 \msolyr.

Galaxy {\it d} has been observed by Pierini et al. (2003), who quote a
galaxy luminosity at 1.4GHz log(L$_{1.4Ghz}$) = 21.64 $\pm$ 0.18,
and a detection at 170$\mu$m.   

Our \hi\ data prove that there is at least one more
galaxy with accordant redshift, the spiral galaxy ESO\,296-G\,026, named galaxy {\it e}.
This galaxy is also strongly forming stars,  with a SFR of almost 
2 \msolyr\.\newline 
Galaxies {\it c} and {\it e} are strong FIR emitters between 25 and
100 $\mu$m (NED).

As already discussed in Section 3.1, the discovery of so many galaxies close to this
group made an assesment of its true nature very difficult. We probed
the entire ATCA field of view, 40$\arcmin$, for \hi\ emission, but found no other
emitter other than galaxies {\it c} and {\it e}. The disturbed distribution of \hi\ emission 
in galaxy {\it e},
together with the equally disturbed distribution of galaxy {\it c }, led
us to believe that galaxy {\it e} is interacting with the group and
we decided to include it in the number of member galaxies.\newline 
The estimate of the galaxy
number density at different radii from the optical centre of the group
shows that SCG0122-3819 is one of the low density groups and that the density
decreases already by one order of magnitude by enlarging the radius to
15$\arcmin$. The strong drop
in galaxy number density outside the original group circle makes this group 
very similar to SCG2159-3210: a central
concentration of galaxies, formed by the original four targets discovered
in the SCGs catalog, plus a diffuse halo of other galaxies.

Two more faint \hi\ emission features are detected
at $\alpha,\delta(2000)$ = 01:25:22, --38:10 and $\alpha,\delta(2000)$ = 01:24:27,
--38:04:30 respectively. These two emission features are observed at the locations
of two optically faint galaxies, possibly dwarf systems. However optical spectroscopic observations
put the redshift of these two galaxies at z $\sim$ 0.08, confirming them
as members of Abell 2911, rather than the group. 
We might hint that at least for one \hi\ detection we are looking at intragroup \hi\ gas
not related to any specific galaxy.
Table~\ref{scg_hi}\ lists quantities derived from the \hi\
line observations.

SCG0122-3819 is another group with a relatively small velocity
dispersion. The only galaxy whose velocity differs from those of
the others by several hundred \kms\ is {\it c}, which shows signs of a 
strong ongoing interaction.
The velocity fields of both {\it c} (=NGC\,546), and {\it e} (=ESO\,296-G\,026;
Fig.~\ref{fig:scgvel_0122}) indicate that the receding side in both
galaxies is the north-eastern half. 

\onlfig{8}{
\begin{figure*}
\centering
\subfigure[]
{\includegraphics[width=9cm]{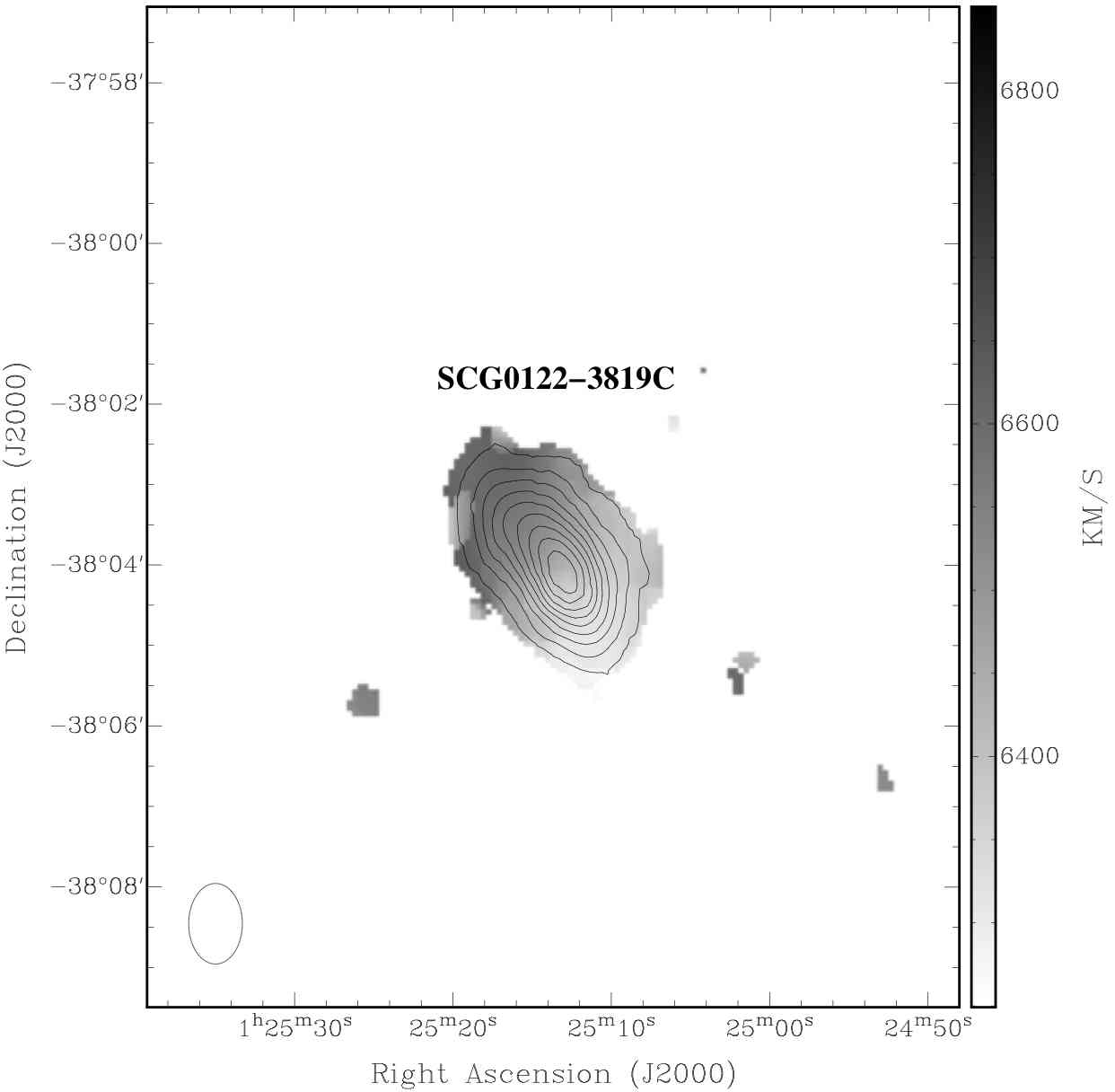}}
\hspace{0.2cm}
\subfigure[]
{\includegraphics[width=9cm]{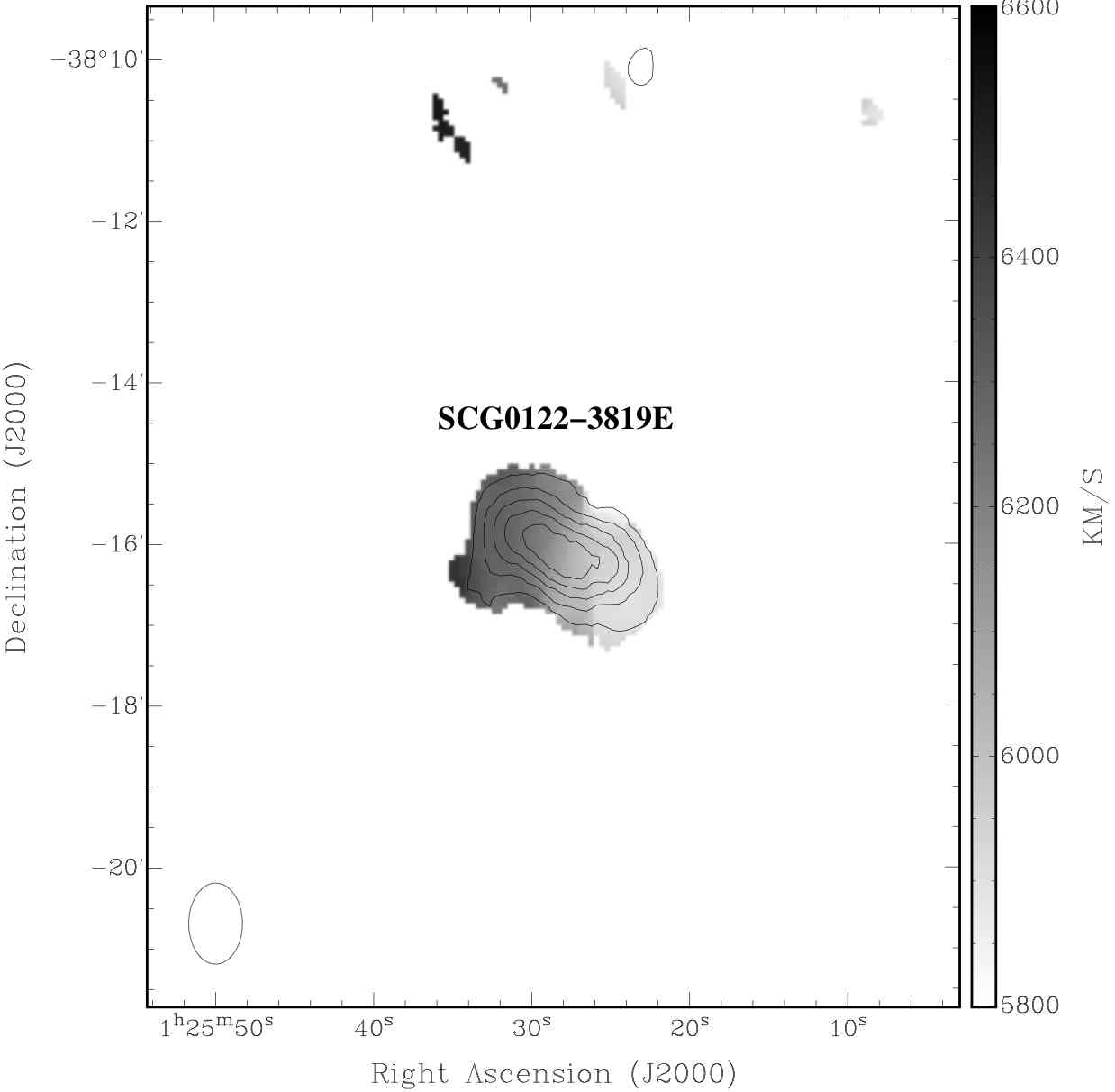}}
\caption{ATCA \hi\ image of SCG\,0122-3819C (left panel) and E (right panel) superimposed on the
\hi\ velocity field. The contours are at the same levels as in
Figure~\ref{fig:scg11hi}; the field of view is  $9\farcm6
\times 10\farcm0$ and the angular resolution ({\it FWHM}= 60$\arcsec
\times$40$\arcsec$) is indicated in the lower left. The velocity scale
is shown on the right, and the orientation of the image is N up and E
to the left.}
\label{fig:scgvel_0122}
\end{figure*}
\afterpage{\clearpage}
}

Both galaxies exhibit signs of tidal
disturbances, not only in the gas distribution, but also its
kinematics. Since both are viewed close to edge-on, their
rotational amplitudes can be used to determine that they are
normal-sized to massive spirals.

\item{}{\bf SCG0141-3429}: this group has five previously detected member galaxies,
of which four were also found in \hi\ emission. The whole group is composed
of lenticulars and spirals, two of which, {\it b} (=IC1722) and {\it c}, have strongly disturbed morphologies,
with a displaced nucleus and bright knots of star formation. Residual
maps in Figure 1 clearly show the irregular distribution of bright
knots in galaxy {\it c}, which are likely star forming regions.
Both galaxies are detected by IRAS as strong infrared emitters.

Galaxies {\it a} and {\it d} are normal spirals, while {\it e}
is a late type galaxy with many SF regions. 
Galaxies {\it b} and {\it c} show strong emission lines,
H$\alpha$, [NII], [SII] [OIII] and [OII], while {\it e}
has very faint emission and off-centre with respect to
the slit position, which was centered on the galaxy nucleus.

Surface photometry for galaxies {\it b} and {\it c} was not
straightforward, due to the irregular morphology of
the targets and due to the bright SF knot visible in
the northern part of galaxy {\it c}.
However good results were obtained after some trial and error
and the best fitting parameters are reported in Table~\ref{scg_pho}. 

The \hi\ emission properties of the four detected galaxies
vary widely, see Figure~\ref{fig:scg0141} and Table~\ref{scg_hi}

\onlfig{9}{
\begin{figure*}
%\hspace*{-20mm}
\resizebox{\hsize}{!}{\includegraphics{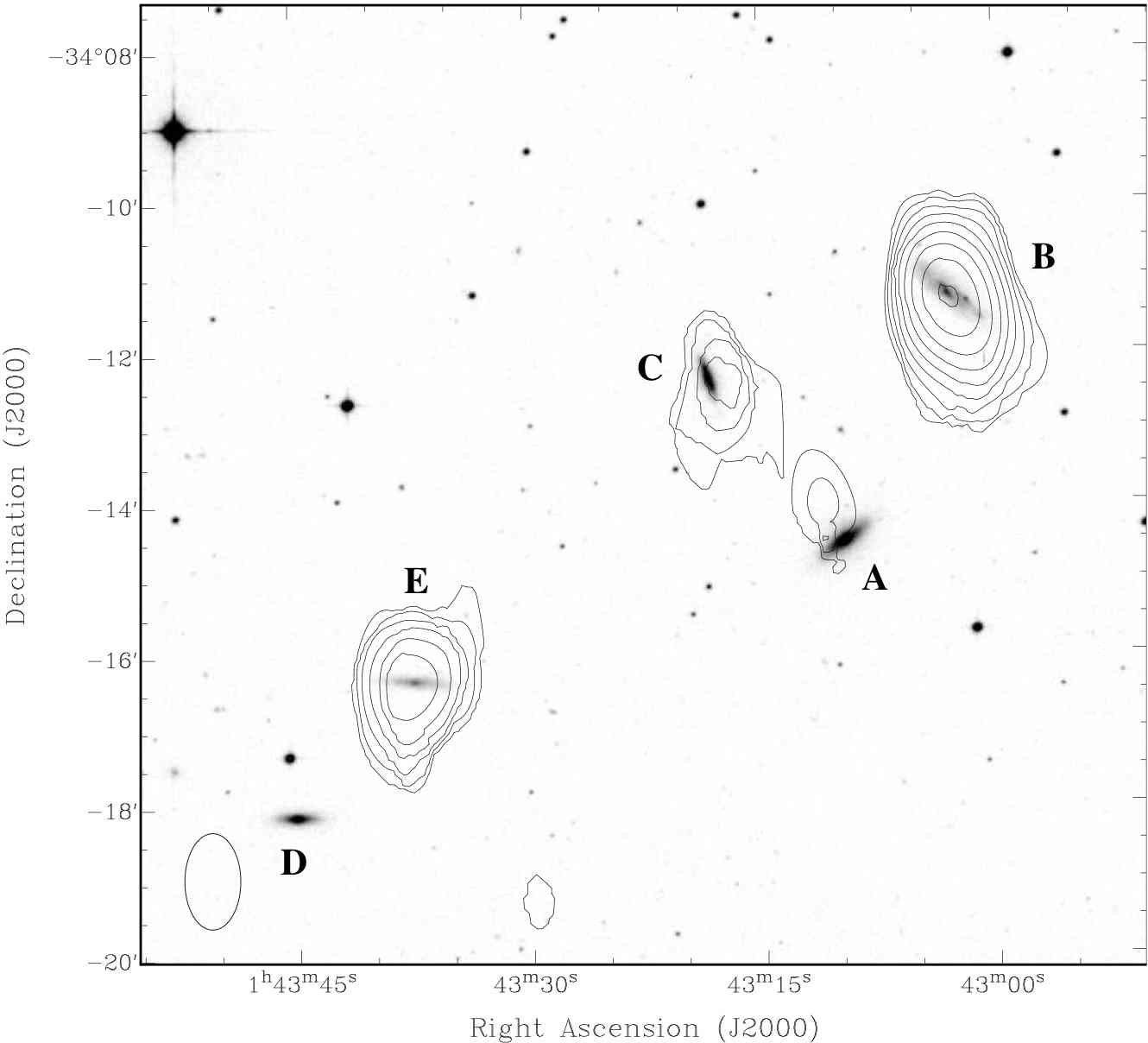}}
\caption{ATCA \hi\ image of SCG\,0141-3429, superimposed on a red DSS-2
optical plate. The contour levels are at 0.17 0.24 0.34 0.48 0.68 0.96
1.36 1.92 2.6 Jy km s$^{-1}$ beam$^{-1}$. The field of view is
$13\farcm4\times 12\farcm7$ and the angular resolution ({\it
FWHM}=$80\arcsec \times 44\arcsec$) is indicated in the lower left.
N is up and E to the left}
\label{fig:scg0141}
\end{figure*}
\afterpage{\clearpage}
}

With the angular resolution of the current data it is not
entirely clear whether the \hi\ gas distribution in galaxy {\it b}
is disturbed or not.
Galaxy {\it e} does appear to have a tidal tail emanating from
the eastern edge of its disk towards south.
Although detected only with low signal-to-noise ratios, the \hi\
gas distributions in galaxies {\it a} and {\it c} are
clearly disturbed. Considering that galaxy {\it a} is the optically
brightest galaxy, it has very little \hi\ gas left, which is
completely displaced with respect to its optical body
along its minor axis, towards {\it c}. 

Radio data confirm that galaxy {\it c} is actively forming stars at a rate
of almost 3 \msolyr, see Table~\ref{scg_rc}.
The \hi\ gas distribution of galaxy {\it c}, in turn, is highly
asymmetric, with a significant fraction of the gas found in
the intergalactic space south of the optical disk, roughly
towards {\it a}.
Only galaxy {\it d}, although confirmed from optical spectroscopy to have a
recession velocity similar to the other group members, is not detected
in \hi\ at all. Note that, although listed as an IRAS far-infrared
source, galaxy {\it d} is not detected in 1.34 GHz radio
continuum emission either (Table~\ref{scg_rc}).

Quantitative HI measurements for all member galaxies are listed in Table~\ref{scg_hi}.

The velocity dispersion of the member galaxies of SCG\,0141-3429
is small, with four galaxies involved in tidal interactions at the
same time, which makes it similar to the Grus quartet of galaxies
(SCG2315-4241, cf. Dahlem 2005).
Despite the strongly disturbed \hi\ gas distributions, the
velocity fields (not displayed here) are still quite regular,
indicating ordered, large-scale gas flows, as it is shown in Figure 10.

\onlfig{10}{
\begin{figure*}
%\hspace*{-20mm}
\resizebox{\hsize}{!}{\includegraphics{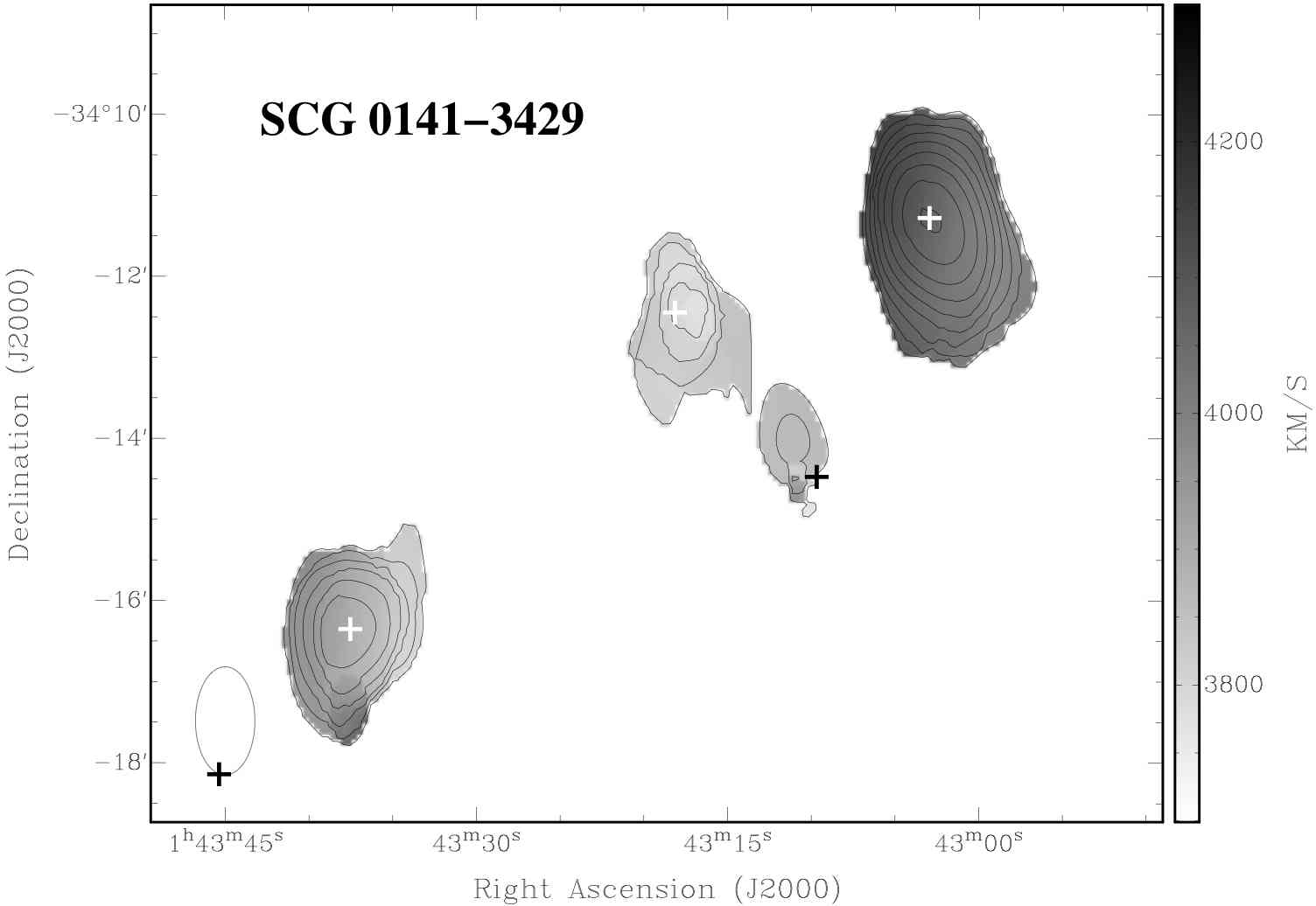}}
\caption{ATCA \hi\ contours of SCG\,0141-3429 superimposed on the
\hi\ velocity field; the crosses indicate the centers of the member
galaxies. The contours are the same as in
Fig.~\ref{fig:scg0141}, the field of view is $12\farcm6\times
10\arcmin1$ and the angular resolution ({\it FWHM}= 38$\arcsec
\times$22$\arcsec$) is indicated in the lower left. The velocity
scale is shown on the right and the orientation is N up and E to the
left.}
\label{fig:scg0141hi_vel}
\end{figure*}
\afterpage{\clearpage}
}

\item{}{\bf SCG0227-4312}: all group members are spiral galaxies and, 
with a total of six optically identified concordant members,
SCG\,0227-4312 is the richest compact group presented here. The member galaxies
show no signs of optical disturbances, but three galaxies, {\it c}, {\it d} and
{\it f} show either a bar or a lens. Galaxy {\it c} is the only one which
also show prominent spiral arms which close around the external ring at the
end of the bar at $\sim$ 33.5$\arcsec$, see Table~\ref{scg_pho}.
All six members of SCG\,0227-4312 can be seen in Fig.~\ref{fig:scg13hi}.

\onlfig{11}{
\begin{figure*}
%\hspace*{-20mm}
\resizebox{\hsize}{!}{\includegraphics{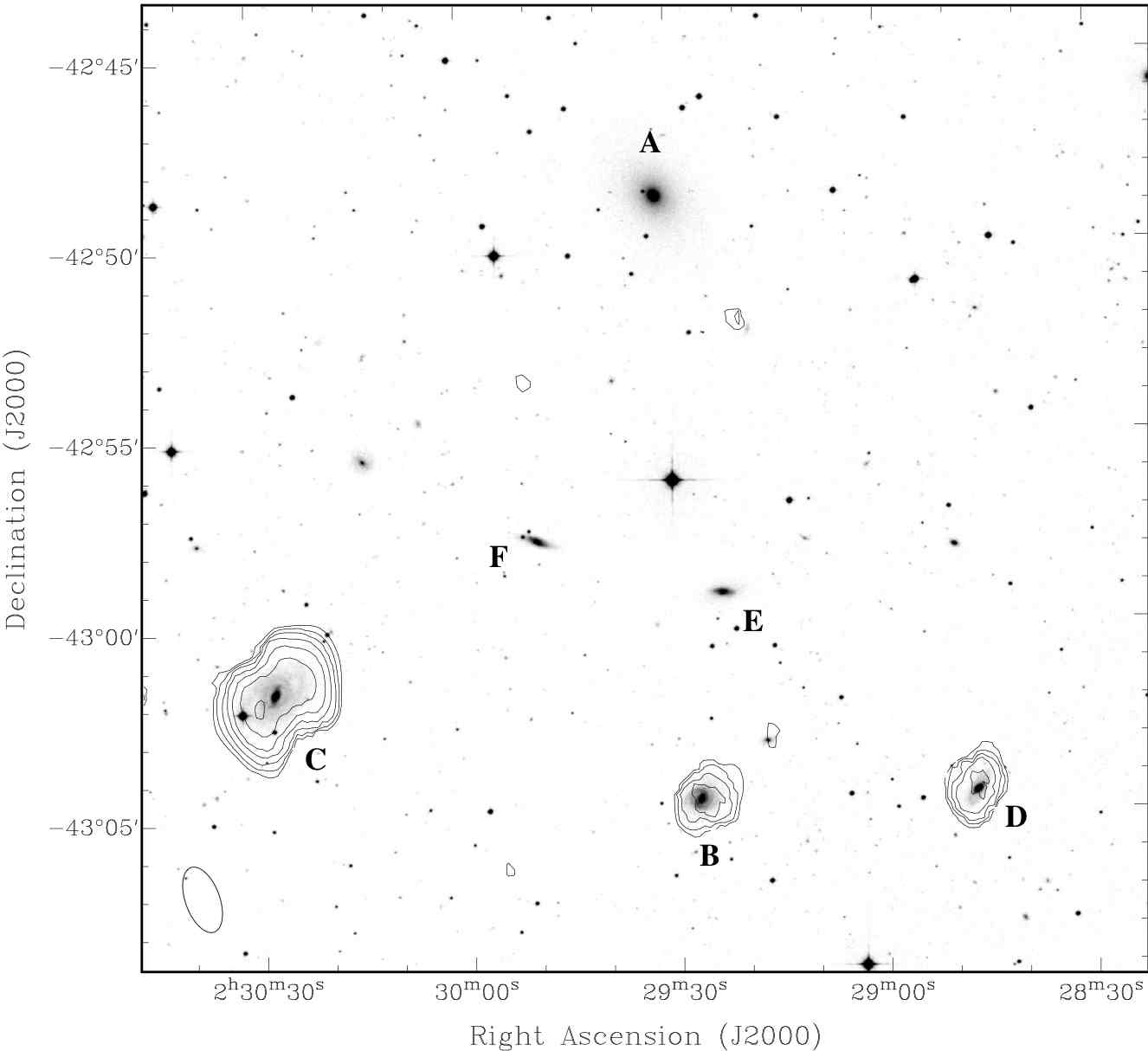}}
\caption{ATCA \hi\ image of SCG\,0227-4312, superimposed on a red DSS-2
optical plate. The contour levels are at 0.2 0.28 0.4 0.56 0.8 1.12 1.6
2.24 Jy km s$^{-1}$ beam$^{-1}$. The field of view is
$26\farcm6\times 25\farcm4$ and the angular resolution ({\it
FWHM}=$38\arcsec \times 22\arcsec$) is indicated in the lower left.
N is up and E to the left}
\label{fig:scg13hi}
\end{figure*}
\afterpage{\clearpage}
}

Of these, only three are detected in \hi\ line emission, namely
{\it b}, {\it c} and {\it d}. 
With the angular resolution of the data presented here no
large-scale irregularities are visible in the \hi\ gas
distributions of these three galaxies, while all show signs
of small disturbances.
The early-type galaxy {\it a} shows no \hi\ line emission
at all and, more surprisingly, neither do the 
edge-on spirals {\it e} and {\it f}, see
Table~\ref{scg_hi}\ for quantitative results.

Only one galaxy, {\it e}, which is classified as a
star forming galaxy, has a detectable radio continuum flux density,
indicating a moderate SFR of around 1 \msolyr, while all
others are very faint at 1.34 GHz.

Both the projected distances and the velocity dispersion of
SCG\,0227-4312 (281 \kms) are large compared to other groups
presented here, see Fig.~\ref{fig:scg0227_vel} and
Table~\ref{scg_hi}. 

\onlfig{12}{
\begin{figure*}
%\hspace*{-20mm}
\resizebox{\hsize}{!}{\includegraphics{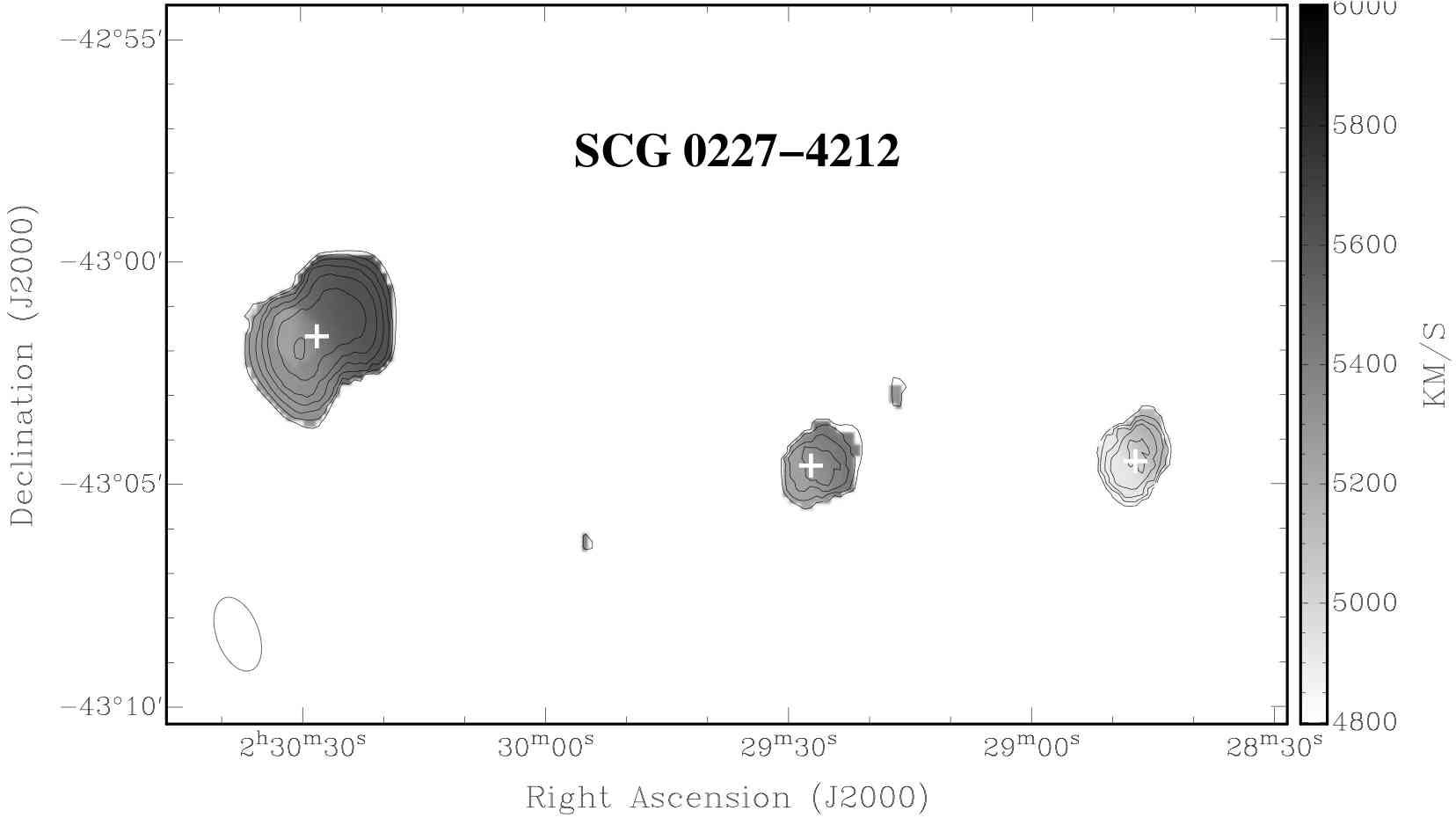}}
\caption{ATCA \hi\ image of SCG\,0227-4212, superimposed on the
\hi\ velocity field. The contours are the same as in
Fig.~\ref{fig:scg13hi}, the field of view is $23\farcm5\times
16\arcmin2$ and the angular resolution ({\it FWHM}= 104$\arcsec
\times$57$\arcsec$) is indicated in the lower left. The velocity
scale is shown on the right and the orientation is N up and E to the
left.}
\label{fig:scg0227_vel}
\end{figure*}
\afterpage{\clearpage}
}

This may explain why this group shows
no direct signs of ongoing interactions. The velocity fields of
the three galaxies showing \hi\ emission are roughly regular.
Note that, despite the wide spread in total \hi\ gas masses,
the rotational amplitudes of the three galaxies detected in
\hi\ emission are quite similar (Table~\ref{scg_hi}).
For similar inclinations, this would indicate comparable
total masses.
\end{itemize}

\section{Discussion}

With the information coming from our optical and radio pilot
sample, we can now ask the following questions:
\begin{itemize}
\item{} how does the population of our SCGs compare with HCGs? 
\item{} do we see any special feature in the intragroup medium?
\item{} what can we say about group evolution?
\end{itemize}

Before answering the first question, we must point out that, being
a pilot study, the number of groups presented here is not
representative of the complete SCGs catalog and definitely not
enough for any statistical conclusion. It is interesting however
to have an idea of what kind of galaxies we are looking at with 
respect to the average properties of compact group galaxies.

In terms of morphological classification, we have that
16 galaxies out of 28, i.e. 60$\%$, are Sa or later, while the others are either E or
S0. Nine out of 28 galaxies in the groups are highly irregular, with signs
of strong morphological disturbances, which corresponds to $\sim$ 30$\%$
of the total number of galaxies. Eleven galaxies show strong bars, which
could be fitted with a Sersic profile, often with a pronounced external
ring.

AGN activity, either in the form of a LINER or an AGN, has been found in seven galaxies,
i.e. about $\sim$ 25$\%$ of our sample, in reasonable agreement with
Coziol et al.(2004); we should point out that the classification of different types 
of nuclear activity, as entered into the {\it Remarks} column of  Table 3,
was revised compared to those published by Coziol et al. (2000).
For the revised classification of galaxies in SCG\,0018-4854 see
Tanvuia (2005), while the classification for SCG0141-3429 is
presented here for the first time.  

About 10 galaxies, $\sim$ 35$\%$, show measurable star formation in the optical, 
but only one, SCG0018-4854A, is a starburst. On the other hand, using
radio continuum data, and assuming as a threshold for starburst
a SFR of 2\msolyr, we find that 6 galaxies out of 28
can be defined star forming galaxies, and half of these six are starbursts.
The difference between the number of star-forming and starburst galaxies
identified in the optical vs. radio regime arises from the fact that we
determine SFRs in different ways. Optical SFRs come from long-slit
spectra probing only the central regions of the galaxies, in order to determine
the dominant type of nuclear activity, if any. On the other hand, radio
SFRs were calculated based on integral radio continuum flux densities
summed up over the surface areas of entire galaxies. These are therefore
global SFRs, rather than measures of nuclear activity.

From these results, we can preliminarily conclude that the SCGs presented
here are on average richer
in spiral members than HCGs; however more data should be taken
into consideration before considering this a firm conclusion. 
SCGs show an higher level of central star
formation, but are comparable with HCGs in terms of occurrence of nuclear 
activity (i.e. LINER and AGNs). We suggest that this confirms the
hypothesis that SCGs are on average younger than HCGs.
This is likely to be due both to the different surveys from which the
two samples were selected and to the different selection criteria
used for the SCGs and HCGs.
Hickson's groups were selected on red plates, while the SCGs catalog has 
been defined through the COSMOS survey, i.e. blue plates,
likely to highlight younger galaxies. In addition to this, the automatic group 
identification used to find SCGs, opposed to the eyeball search for HCGs, has eliminated
the bias of selecting preferentially higher surface brightness groups and
groups with two or more galaxies of comparable magnitude, which is present in the
HCGs (Prandoni et al., 1994).

The sensitivity of \hi\ observations is about $2\times 10^8$ \msol;
however twelve (43$\%$) of the optically identified 28 member galaxies in the
six groups are not detected in \hi\ line emission. Of these 12,
8 galaxies are classified as early-type (E or S0; SCG2159-3210{\it b} and {\it c},
SCG2353-6101{\it c}, SCG0122-3819{\it b} and {\it d}, SCG0141-3429{\it d}
and SCG0227-4312{\it a} and {\it e}). 

On the other hand, 4 galaxies without \hi\ detections are later types in which one
would normally expect to find atomic hydrogen gas. While for 
SCG2159-3210{\it d} and SCG0122-3819{\it a} this can be
explained by stripping caused by the  strong interaction these galaxies are 
experiencing, it is very difficult to understand what has happened to galaxy
SCG0227-4312{\it f}, which seems completely undisturbed and should not have
such \hi\ deficiency.
A search of the Chandra and XMM archives returned no result on this galaxy,
so at the moment we are left with an unsolved riddle.

This lack of gas is confirmed by the fact that we find an average
\hi\ deficiency of 0.61. We would like to stress that this deficiency
is an upper limit to the real estimate of \hi\ deficiency in
groups, as we did not apply any correction for \hi\ absorption
for the presence of an active nucleus or thick gas.

Tidal tails are more evident in \hi\ observations than in the optical,
with 9 galaxies showing such feature in the \hi\ gas distributions, while 
only three have optical tidal tails. Other signs of morphological disturbances can be
detected in another 4 galaxies, bringing the total of interacting galaxies
to 13, almost half of the sample.
From these results, we can conclude that \hi\ is a very powerful tool to
trace the effects of the interactions among the member galaxies
out to a large distance from the centre of the galaxies themselves.

\hi\ data give more information on the dynamical evolutionary state of the groups:
following the  classification scheme by VM01, we
can identify which phase the six groups presented here are currently in, from
the so called {\it phase 1}, with the gas mainly centered on the member galaxies,
to {\it phase 2}, with extended tidal tails and perturbed gas distribution to
{\it phase 3a}, with almost no or very little \hi\ left, or {\it phase 3b},
with all the \hi\ gas distributed in the group potential well and not
centered on any single galaxy.

SCG\,2159-3210 is at a late stage of phase 3a in its dynamical
evolution. No \hi\ is found in the three strongly interacting
galaxies in this group and the \hi\ gas in galaxy {\it a} exhibits
an asymmetric distribution and a tidal tail. This group is
extremely deficient in \hi\ gas, because almost none was
detected in its member galaxies. The evolutionary stage finds
further confirmation in the group-wide X-ray emission (Ponman et al., 1996).

SCG\,2353-6101, on the other hand, must still be in an early
stage of its dynamical evolution, because there are only few
indications that the \hi\ gas in its galaxies is disturbed.

The four galaxies near the centre of SCG\,0018-4854 have a
common \hi\ envelope, which indicates that they are in phase 3b
according to VM01. Galaxy {\it e}, farther
away from the central four, is not part of the ongoing interactions
and it is located outside their common envelope.

Although both SCG\,0122-3819{\it c} and SCG\,0122-3819{\it e} show signs of interactions
in the form of tidal tails, these are not very pronounced yet
and their \hi\ gas is still well aligned with their stellar
disks. Therefore, this group is most likely still in phase 1
of its dynamical evolution, possibly just starting to move into phase 2.

SCG\,0141-3429 is far more developed than SCG\,0122-3819. The
two galaxies {\it a} and {\it c} have lost, or are in the process of losing,
most of their \hi\ gas into the intergalactic space, with
asymmetric \hi\ distribution and tidal tails. This would place this group in
phase 2, having galaxies {\it b} and {\it e} at
an earlier stage of their development, with \hi\ that is
slightly disturbed, but still tightly bound to the stellar
disks.

SCG\,0227-4312 is still dynamically young, in phase 1 of its
evolution. The late-type galaxies show little, if any, sign of
gravitational disturbance in their \hi\ distributions and
kinematics. The only galaxies that have no \hi\ are early-type
galaxies, with the exception of galaxy {\it f}. According to Casasola 
et al. (2004) this is further evidence that these galaxies did not 
experience any interaction at all.

From the above it is clear that a classification of each
group into a specific evolutionary stage is not easy. Nevertheless,
the data show that some groups are still in early stages of their
evolution, while others are already quite evolved.
It should be noted that the group in phase 3b does not
show \hi\ deficiency; the same is true for the similarly classified
groups in VM01, see for example HCG49.
Another interesting point is that both Sy2 in our sample belong
to groups in the last evolutionary phase. However, while 
SCG2159-3210{\it a} is located outside the main site of interaction
and it has retained some of its \hi\ gas, SCG0018-4854{\it b}
is sitting in the middle of an interacting group and it is completely
devoid of \hi.

The different \hi\ gas distributions observed in the SCGs presented
here highlight the importance of \hi\ gas as diagnostic for studies of  the
dynamical evolution of galaxy groups.

\section{Conclusions}

In this pilot study of six compact groups drawn from the
Southern Compact Groups catalog we have found that;
\begin{itemize}
\item{} Four groups are real compact groups, while two are
{\it core+halo} systems.
\item{} The three dimensional velocity dispersion and mass of our groups
are very similar to HCGs; the range covered in crossing 
time and M/L is compatible with the distribution 
observed for HCGs. 
\item{} Out of 28 galaxes, 35$\%$ of them are forming stars in their
central region, and approximately
60$\%$ of them are spirals of morphological type Sa or later.
\item{} Six galaxies, about 20$\%$ of the total, are forming stars throughout
the whole galaxy and three of them are starbursts.
\item{} Half of the galaxies show signs of interactions.
\item{} The \hi\ distribution in our groups approximately reveal
the presence of all the four evolutionary stages highlighted by
VM01. The average \hi\ deficiency is
in agreement with what has been measured for galaxes in VM01.
\end{itemize}
More detailed studies of individual groups will be conducted by
us, based on multi-wavelength observations. At the same time, we
will be collecting a more complete database of all groups in
our statistically complete sample comprising a total of 50 SCGs,
whose brightest galaxy has b$_{j} \le$ 14.5, to increase the statistical significance of our results.

The sample of SCGs presented here is too small to perform
detailed comparisons with other, larger samples. General trends
found in larger samples, such as enhanced levels of nuclear
activity in interacting galaxies and \hi\ deficiencies in groups
at late stages of their dynamical evolution, are tentatively
corroborated. However, firmer conclusions will have to await the
availability of a larger number of suitable data sets.
As part of this, more \hi\ observations of SCGs will be carried
out and presented at a later stage.

\acknowledgements{
MD gratefully acknowledges the kind hospitality and travel grant
received from ESO during a visit to Santiago de Chile, where part
of the work presented here was carried out.
%
%This research has made use of the NASA Extragalactic Database (NED),
%whose contributions to this paper are gratefully acknowledged.
%
The Digitized Sky Survey was produced at the Space Telescope
Science Institute under U.S. Government grant NAG W-2166. The
National Geographic Society -- Palomar Observatory Sky Atlas
(POSS-I) was made by the California Institute of Technology
with grants from the National Geographic Society.
}

\setcounter{table}{7}
\begin{table*}
\caption{\label{scg_hi} HI results for the individual group galaxies.}
\begin{tabular}{lccccccc}
\hline
Name of the galaxy & Other name & $v$(opt)$_{\rm hel}^{\rm b}$ & $v$(HI)$_{\rm hel}^{\rm c}$ & $v$(HI)$_{\rm max}^{\rm c}$ & {\it f(HI)} &  {\it M(HI)} &Comments \\ 
                   &            & (km s$^{-1}$)              & (km s$^{-1}$)             & (km s$^{-1}$)              &  (Jy km s$^{-1}$) & ($10^9$ \msol) \\ \hline  
SCG2159-3120A & NGC7172 & 2575$\pm$28$^{\rm a}$ & 2484$\pm$100 & 237$\pm$50 & 0.82  & $>0.19^{\rm d}$ & Sa{\it pec}, Sy2, tidal tail \\
SCG2159-3120B & NGC7176 & 2525$\pm$29$^{\rm a}$ & --- & --- & $<0.13^{\rm e}$ &  $<0.03^{\rm e}$ & E pec: \\
SCG2159-3120C & NGC7173 & 2696$\pm$24$^{\rm a}$ & --- & --- & $<0.13^{\rm e}$ &  $<0.03^{\rm e}$ & E+ pec: \\
SCG2159-3120D & NGC7174 & 2778$\pm$29$^{\rm a}$ & --- & --- & $<0.13^{\rm e}$ &  $<0.03^{\rm e}$ & Sab{\it pec} tidal tail \\
     &           &             &     &     &                &                 &            \\
SCG2353-6101A & ESO111-G010 & 4833$\pm$38 & 4800$\pm$30 & 25$\pm$20 & 0.76 & 0.54 & \\
SCG2353-6101B & ESO111-G009 & 4246$\pm$42 & 4167$\pm$30 & 138$\pm$30 & 4.42 & 3.11 & tidal tail \\
SCG2353-6101C & --- & 4535$\pm$36 & --- & --- & $<0.16^{\rm f}$ & $<0.11^{\rm f}$  & \\
     &           &             &     &     &                &                 &            \\
SCG0018-4854A & NGC92 & 3443$\pm$23 & 3400$\pm$30 & 162$\pm$30 & 8.35 & $>3.29^{\rm d}$ & SF; tidal tail \\
SCG0018-4854B & NGC89 & 3326$\pm$32 & --- & --- & $<0.07^{\rm g}$ & $<0.02^{\rm g}$ & \\
SCG0018-4854C & NGC87 & 3467$\pm$34 & 3405$\pm$20 & 46$\pm$20 & 1.44 & 0.57 & \\
SCG0018-4854D & NGC88 & 3506$\pm$63 & 3413$\pm$20 & 67$\pm$20 & 2.05 & 0.81 & \\
SCG0018-4854E & ESO194-G013 & 3187$\pm$20 & 3210$\pm$20 & 83$\pm$20 & 6.39 & 2.52& \\
     &           &             &     &     &                &                 &            \\
SCG0122-3819A & NGC534 & 5858$\pm$39 & --- & --- & $<0.12^{\rm h}$  & $<0.16^{\rm h}$ & \\
SCG0122-3819B & NGC544 & 5998$\pm$28 & --- & --- & $<0.12^{\rm h}$ & $<0.16^{\rm h}$ & \\
SCG0122-3819C & NGC546 & 6583$\pm$27 & 6414$\pm$30 & 228$\pm$30 & 7.38 & 9.54 & tidal tail \\
SCG0122-3819D & NGC549 & 6060$\pm$30 & --- & --- & $<0.12^{\rm h}$ & $<0.16^{\rm i}$ & \\
SCG0122-3819E & ESO296-G026 & 6176$\pm$10$^{d}$ & 6104$\pm$30 & 266$\pm$30 & 4.59 & 5.93 & tidal tail, Sab \\
SCG0122-3819F &    ---   & --- & 6020$\pm$20 & 25$\pm$20 & 0.27 & 0.35 & optical counterpart \\
              &          &     &             &           &      &      & not identified\\
SCG0122-3819G &    ---   & --- & 5745$\pm$20 & 40$\pm$20 & 0.15 & 0.20 & optical counterpart\\
              &          &     &             &           &      &      & not identified\\
              &          &     &             &           &      &      &            \\
SCG0141-3429A & IC1724 & 3829$\pm$70 & 3855$\pm$30 & 33$\pm$20 & 0.33 & 0.17 & displaced HI gas \\
SCG0141-3429B & IC1722 & 4176$\pm$17 & 4074$\pm$30 & 114$\pm$30 & 4.56 & 2.37 & tidal tail? \\
SCG0141-3429C & ESO353-G036 & 3808$\pm$16 & 3749$\pm$40 & 119$\pm$40 & 0.99 & 0.51 & tidal tail \\
SCG0141-3429D & IRASF01415-3433 & 3379$\pm$66 & --- & --- & $<0.12^{\rm l}$ & $<0.06^{\rm l}$ & \\
SCG0141-3429E & ESO353-G039 & 3913$\pm$18 & 3883$\pm$20 & 103$\pm$20 & 1.88 & 0.98 & tidal tail \\
     &           &             &     &     &                &                 &            \\
SCG0227-4312A & IC1812 & 5240$\pm$24 & --- & --- & $<0.12^{\rm c}$ & $<0.12^{\rm c}$ & \\
SCG0227-4312B & IC1810 & 5508$\pm$49 & 5329$\pm$30 & 161$\pm$30 & 0.82 & 0.81 & \\
SCG0227-4312C & ESO246-G021 & 5558$\pm$23 & 5455$\pm$20 & 208$\pm$20 & 4.98 & 4.96 & \\
SCG0227-4312D & ESO246-G015 & 5129$\pm$34 & 5028$\pm$40 & 201$\pm$40 & 0.62 & 0.62 & \\
SCG0227-4312E & ESO246-G016 & 5019$\pm$39 & --- & --- & $<0.12^{\rm l}$ & $<0.12^{\rm l}$ & \\
SCG0227-4312F & ESO246-G020 & 5850$\pm$23 & --- & --- & $<0.12^{\rm l}$ & $<0.12^{\rm l}$ & \\ \hline
\end{tabular}
{\small a) From Hickson et al., (1992).} \newline
{\small b) From optical spectroscopy, unless otherwise specified.}\newline
{\small c) From HI data presented here.} \newline
{\small d) Lower limit because HI gas partly in absorption against nuclear
radio continuum emission.} \newline
{\small e) 5-$\sigma$ upper limit, with a 1-$\sigma$ rms per channel map of
1.0 mJy beam$^{-1}$ and a channel width of 26.4 kms$^{-1}$.}\newline
{\small f) 5-$\sigma$ upper limit, with a 1-$\sigma$ rms per channel map of
1.2 mJy beam$^{-1}$ and a channel width of 26.4 kms$^{-1}$.}\newline
{\small g) 5-$\sigma$ upper limit, with a 1-$\sigma$ rms per channel map of
0.53 mJy beam$^{-1}$ and a channel width of 26.4 kms$^{-1}$.}\newline
{\small h) 5-$\sigma$ upper limit, with a 1-$\sigma$ rms per channel map of
0.19 mJy beam$^{-1}$ and a channel width of 26.4 kms$^{-1}$.}\newline
{\small i) From Mathewson et al. (1992)}\newline
{\small l) 5-$\sigma$ upper limit, with a 1-$\sigma$ rms per channel map of
0.9 mJy beam$^{-1}$ and a channel width of 26.4 kms$^{-1}$.} 
\end{table*}

\afterpage{\clearpage}

\begin{table}
\caption{1.34 GHz continuum emission from members of the six groups presented here}
\label{scg_rc}
\begin{tabular}{ccccc}
\hline
  Galaxy Name & {\it f(1.34)} & $P_{21}^{1.34}$ & $\nu_{\rm SN}$ & {\it  SFR} \\
              &     (mJy)     & (W\,Hz$^{-1}$) & (yr$^{-1}$)    &  [\msolyr] \\\hline
SCG2159-3210A &  38.6   & 5.82 & ---$^{\rm a}$ & ---$^{\rm a}$ \\
SCG2159-3210B & $<0.38^{\rm b}$ & $<0.05$ & $<0.0007$ & $<0.018$ \\
SCG2159-3210C &   1.0   & 0.15 & 0.002 & 0.05 \\
SCG2159-3210D &  18.4   & 2.78 & 0.038 & 0.92 \\
              &         &      &       &      \\
SCG0018-4854A & 177.3  & 44.50 & 0.616 & 14.71 \\
SCG0018-4854B &  12.3  &  3.09 & ---$^{\rm a}$ & ---$^{\rm a}$ \\
SCG0018-4854C & $<0.9^{\rm c}$ & $<0.23$ & $<0.003$ & $<0.07$ \\
SCG0018-4854D &   2.9  &  0.73 & 0.010 & 0.24 \\
SCG0018-4854E &   6.4  &  1.61 & 0.022 & 0.53 \\
              &         &      &       &      \\
SCG0122-3819A & $<0.33^{\rm d}$ & $<0.26$ & $<0.004$ & $<0.09$\\
SCG0122-3819B &   0.45  & 0.37 & ---$^{\rm a}$ & ---$^{\rm a}$\\
SCG0122-3819C &   7.88  & 6.44 & 0.089 & 2.13 \\
SCG0122-3819D & $<0.33^{\rm d}$ & $<0.26$ & $<0.004$ & $<0.09$\\
SCG0122-3819E &   7.31  & 5.98 & 0.083 & 1.98 \\
SCG0122-3819F & $<0.33^{\rm d}$ & $<0.26$ & $<0.004$ & $<0.09$\\
SCG0122-3819G & $<0.33^{\rm d}$ & $<0.26$ & $<0.004$ & $<0.09$\\
              &         &      &       &      \\
SCG0141-3429A & $<1.3^{\rm e}$ & $<0.42$ & $<0.006$ & $<0.14$ \\
SCG0141-3429B &   1.5  & 0.49 & 0.007 & 0.16 \\
SCG0141-3429C &  26.4  & 8.57 & 0.119 & 2.83 \\
SCG0141-3429D & $<1.3^{\rm e}$ & $<0.42$ & $<0.006$ & $<0.14$ \\
SCG0141-3429E & $<1.3^{\rm e}$ & $<0.42$ & $<0.006$ & $<0.14$ \\
              &         &      &       &      \\
SCG0227-4312A & $<1.0^{\rm f}$ & $<0.64$ & $<0.009$ & $<0.20$ \\
SCG0227-4312B & $<1.0^{\rm f}$ & $<0.64$ & $<0.009$ & $<0.20$ \\
SCG0227-4312C & $<1.0^{\rm f}$ & $<0.64$ & $<0.009$ & $<0.20$ \\
SCG0227-4312D & $<1.0^{\rm f}$ & $<0.64$ & ---$^{\rm a}$ & ---$^{\rm a}$ \\
SCG0227-4312E &   5.59 & 3.55 & 0.049 & 1.17 \\
SCG0227-4312F & $<1.0^{\rm f}$ & $<0.62^{\rm f}$ & $<0.009$ & $<0.20$ \\ \hline
\end{tabular}
{\small a) $\nu_{\rm SN}$ and {\it SFR} could not be determined; evidence for\newline
    emission from an active nucleus (see Table~\ref{scg_pho}).} \newline
{\small b) 5-$\sigma$ upper limit for an rms noise of 0.075 mJy beam$^{-1}$.}\newline
{\small c) 5-$\sigma$ upper limit for an rms noise of 0.185 mJy beam$^{-1}$.}\newline
{\small d) 5-$\sigma$ upper limit for an rms noise of 0.065 mJy beam$^{-1}$.}\newline
{\small e) 5-$\sigma$ upper limit for an rms noise of 0.26 mJy beam$^{-1}$.}\newline
{\small f) 5-$\sigma$ upper limit for an rms noise of 0.20 mJy beam$^{-1}$.}
\end{table}

\Online

The next nine figures show the optical images and the GALFIT residual images
for the member galaxies of the groups presented here. The following 10 additional figures
show the results obtained from the radio data, overlapped on the optical images

%% IMAGES FIRST

% optical images of the groups
\setcounter{figure}{0}
\begin{figure*}
\includegraphics{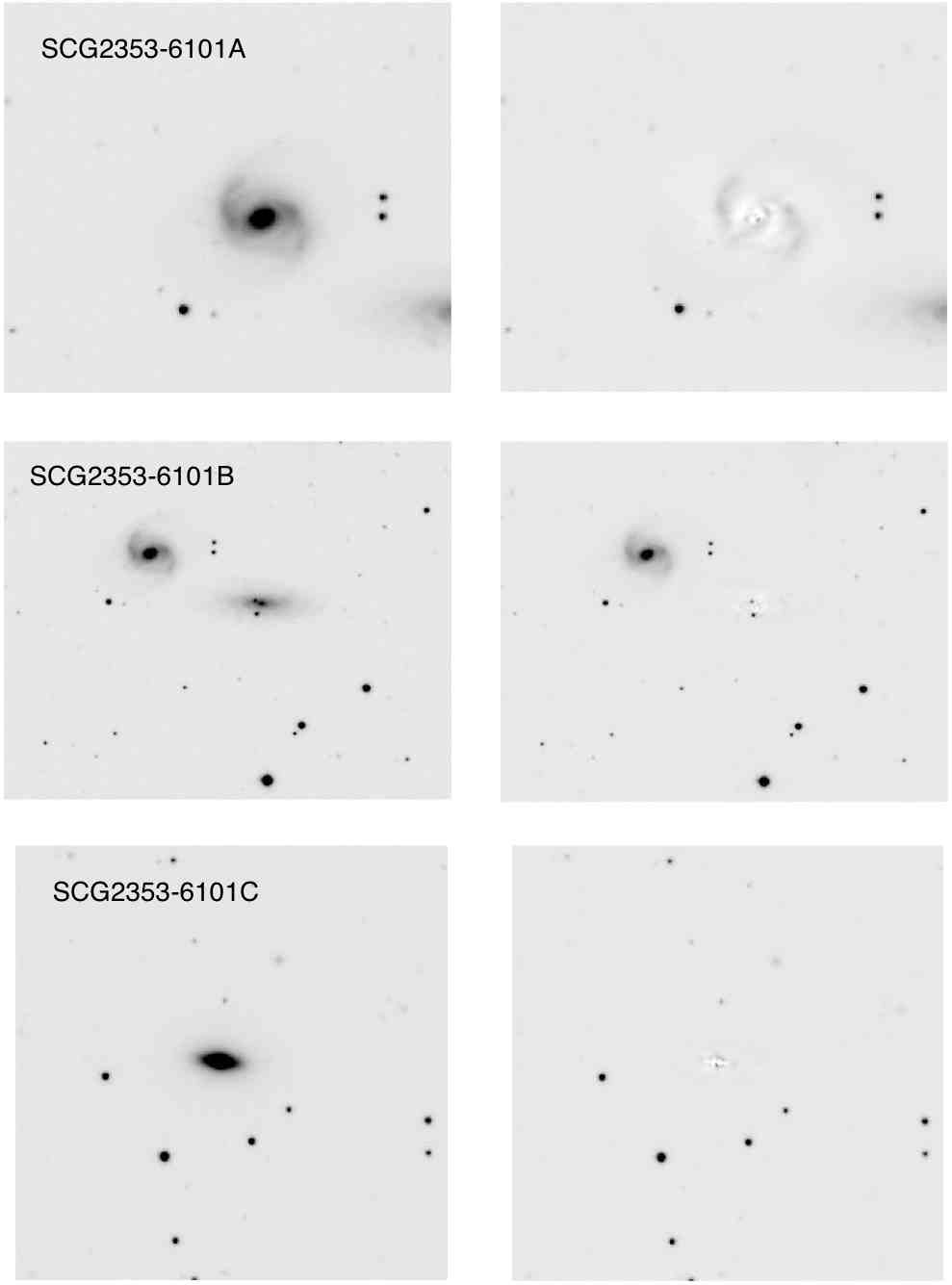}
\caption{Optical images (left panel) and GALFIT residual images (right
panel) for SCG2353-6101, from galaxy {\it a} to {\it c} from top to bottom.
Residual brightness for galaxy A is of the order of 5$\%$ of the
peak brightness, while for the other galaxies is less than 1$\%$. In all images N is
up and E is left.}
\end{figure*}

\afterpage{\clearpage}

\setcounter{figure}{0}
\begin{figure*}
\includegraphics{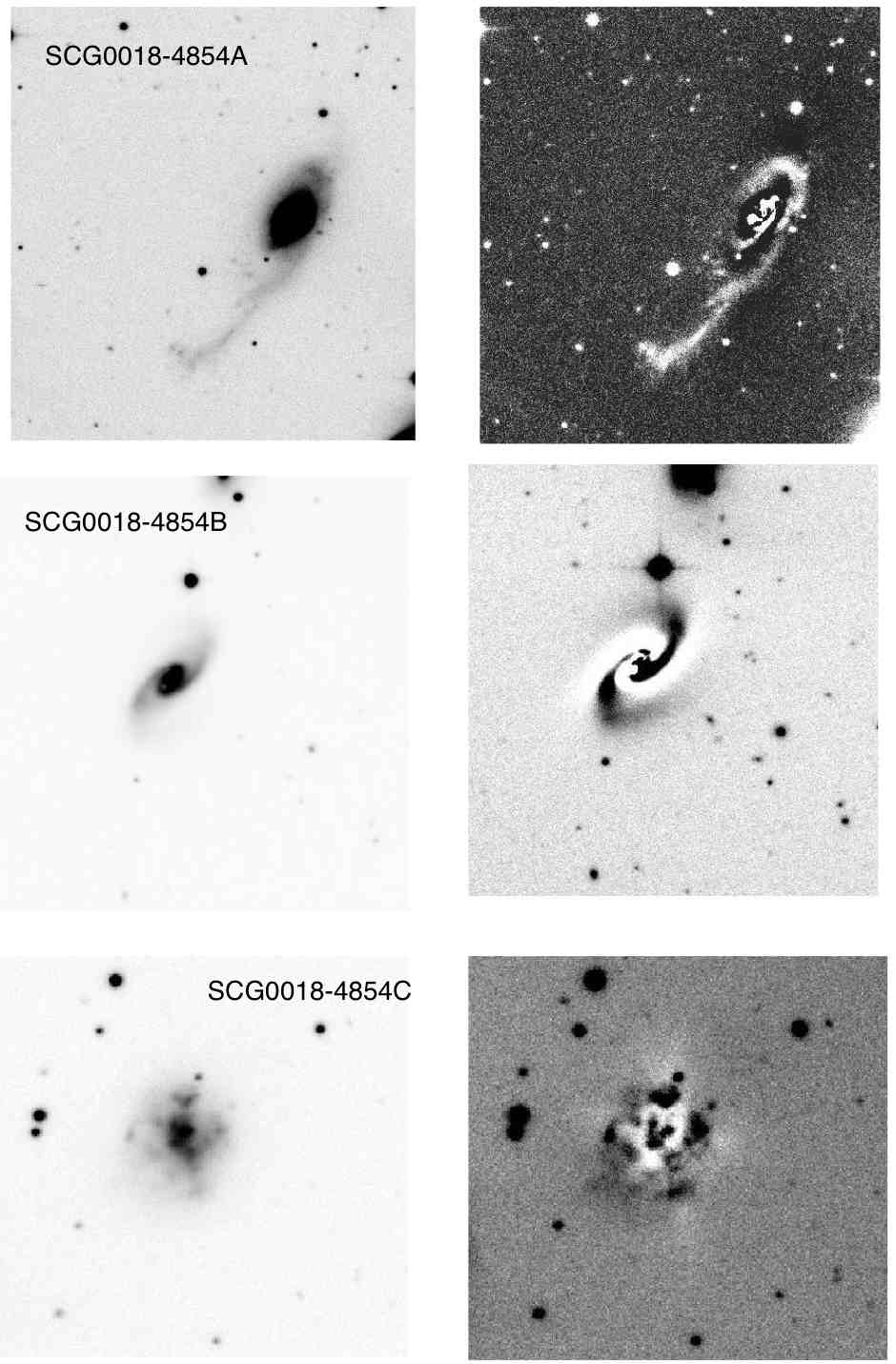}
\caption{ (continued) Optical images (left panel) and GALFIT residual images (right
panel) for SCG0018-4854, from galaxy {\it a} to {\it c} from top to bottom. 
Residual brightness for all galaxies is of the order of 10$\%$ of the
peak brightness. In all images N is
up and E is left.}
\end{figure*}

\afterpage{\clearpage}

\setcounter{figure}{0}
\begin{figure*}
\includegraphics{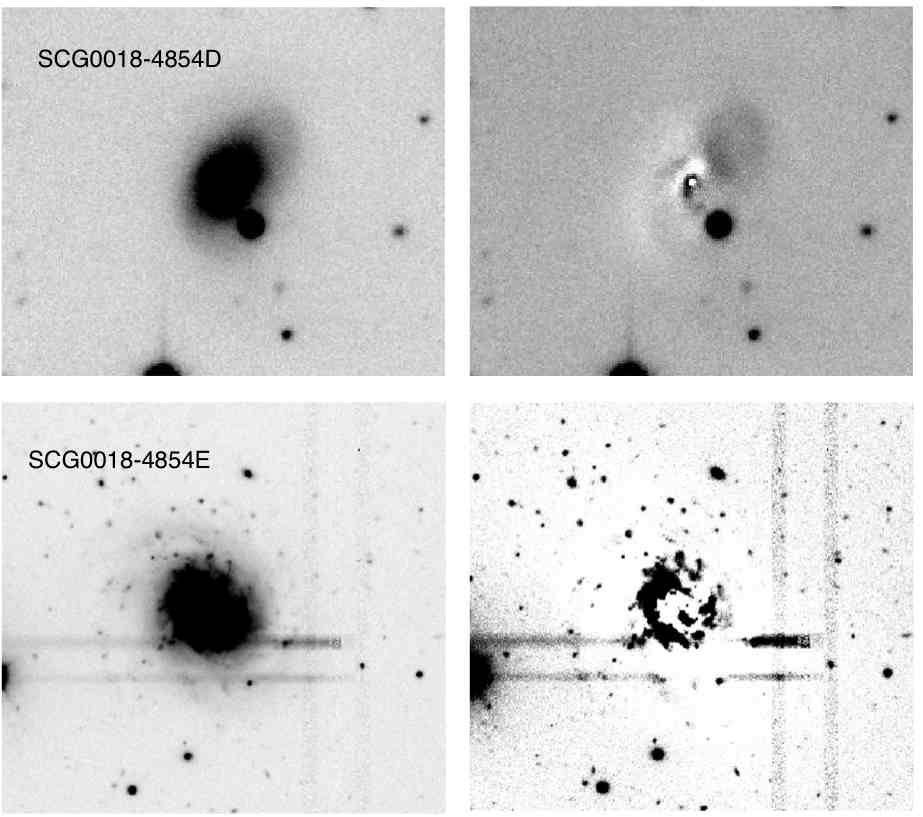}
\caption{ (continued) Optical images (left panel) and GALFIT residual images (right
panel) for SCG0018-4854, from galaxy {\it d} to {\it e} from top to bottom. 
Residual brightness for all galaxies is of the order of 7$\%$ of the
peak brightness.
In all images N is up and E is left.}
\end{figure*}

\afterpage{\clearpage}

\setcounter{figure}{0}
\begin{figure*}
\includegraphics{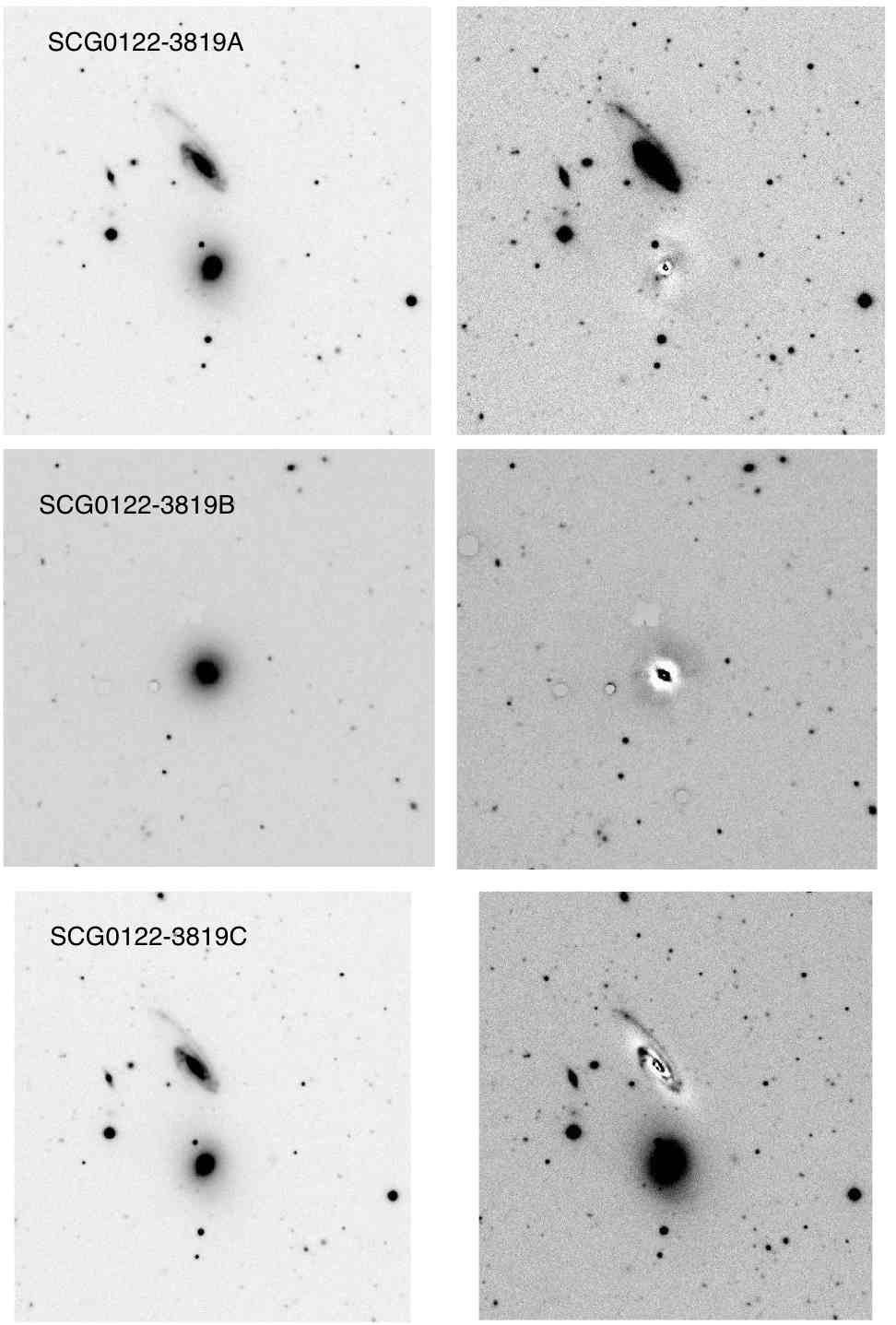}
\caption{ (continued) Optical images (left panel) and GALFIT residual images (right
panel) for SCG0122-3819, from galaxy {\it a} to {\it c} from top to bottom. 
Residual brightness for all galaxies is of the order of 5$\%$ of the
peak brightness, with the exception of galaxy B, where the residuals
are of the order of 15$\%$. In all images N is
up and E is left.}
\end{figure*}

\afterpage{\clearpage}

\setcounter{figure}{0}
\begin{figure*}
\includegraphics{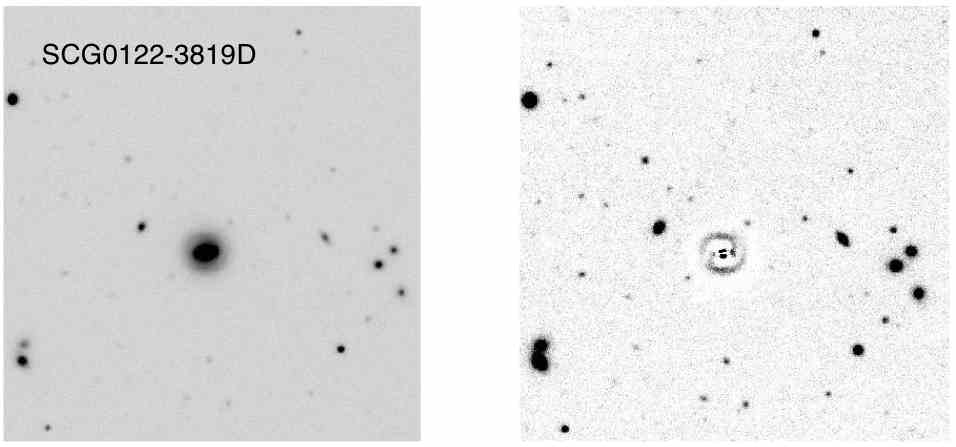}
\caption{ (continued) Optical images (left panel) and GALFIT residual images (right
panel) for SCG0122-3819{\it d}; residuals are of the order of 8$\%$ of
the peak brightness. N is up and E is left.}
\end{figure*}

\afterpage{\clearpage}

\setcounter{figure}{0}
\begin{figure*}
\includegraphics{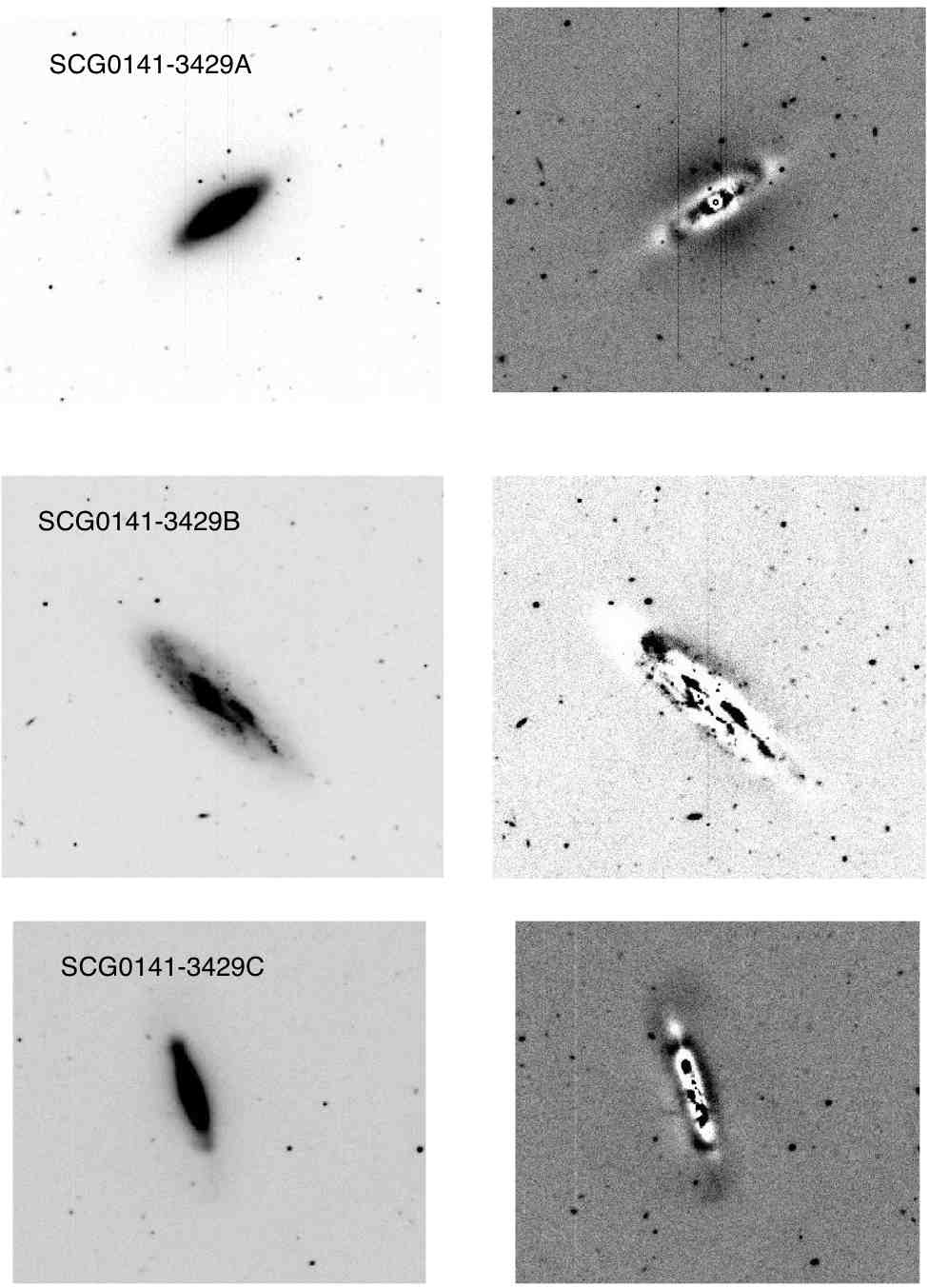}
\caption{ (continued) Optical images (left panel) and GALFIT residual images (right
panel) for SCG0141-3429, from galaxy {\it a} to {\it c} from top to bottom. 
Residual brightness for all galaxies is of the order of 12$\%$ of the
peak brightness. In all images N is
up and E is left.}
\end{figure*}

\afterpage{\clearpage}

\setcounter{figure}{0}
\begin{figure*}
\includegraphics{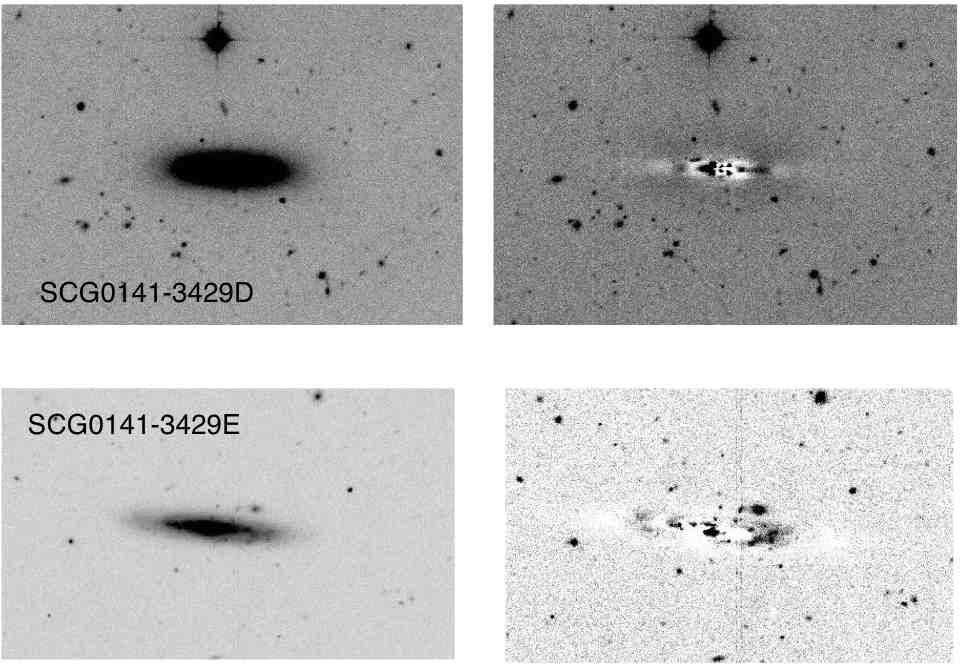}
\caption{ (continued) Optical images (left panel) and GALFIT residual images (right
panel) for SCG0141-3429, from galaxy {\it d} to {\it e} from top to bottom. 
Residual brightness for all galaxies is of the order of 10$\%$ of the
peak brightness. In all images N is
up and E is left.}
\end{figure*}

\afterpage{\clearpage}

\setcounter{figure}{0}
\begin{figure*}
\includegraphics{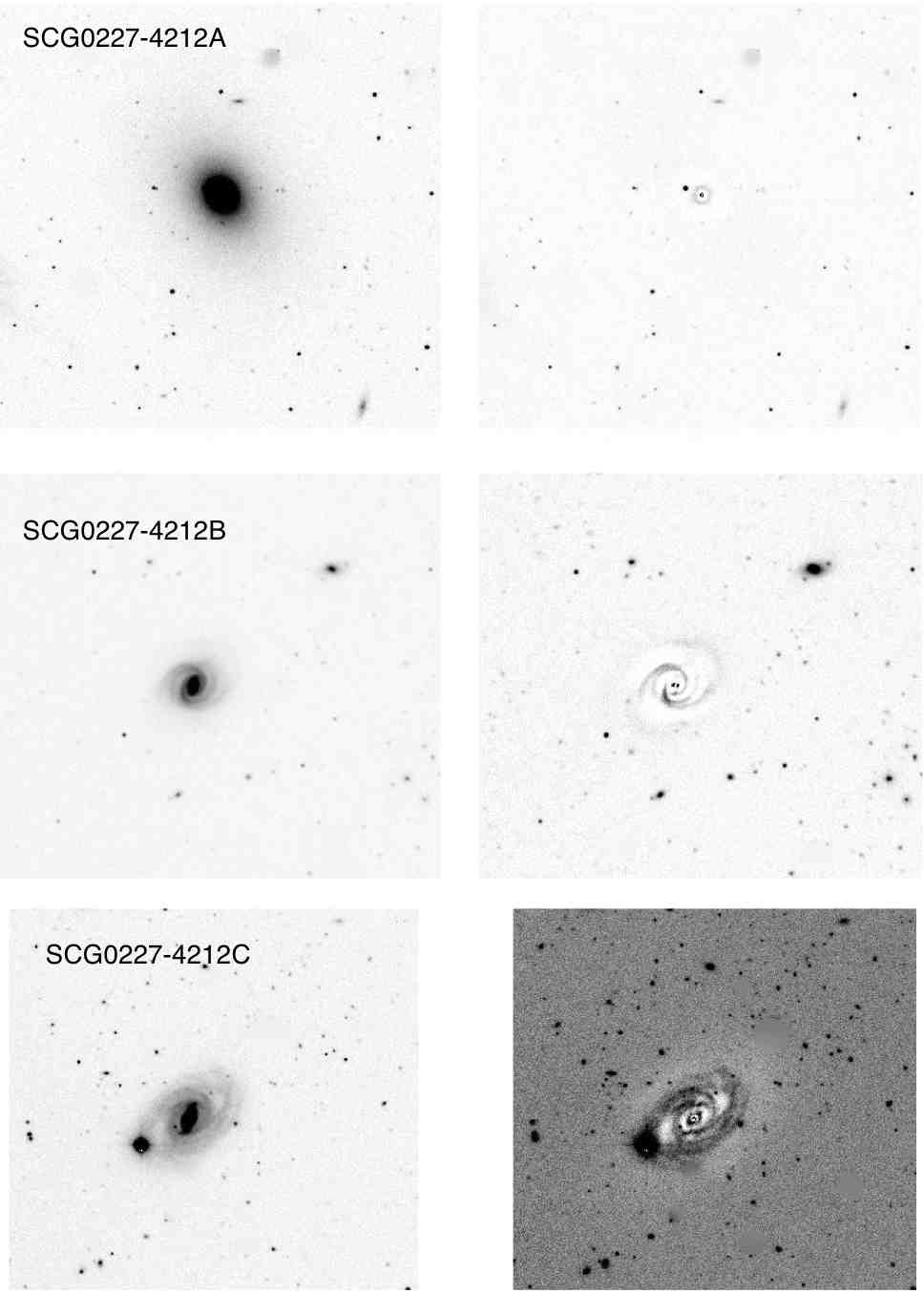}
\caption{ (continued) Optical images (left panel) and GALFIT residual images (right
panel) for SCG0227-4312, from galaxy {\it a} to {\it c} from top to bottom. 
Residual brightness for all galaxies is of the order of 3$\%$ of the
peak brightness, with the exception of galaxy C, where the residuals
are of the order of 8$\%$. In all images N is
up and E is left.}
\end{figure*}

\afterpage{\clearpage}

\setcounter{figure}{0}
\begin{figure*}
\includegraphics{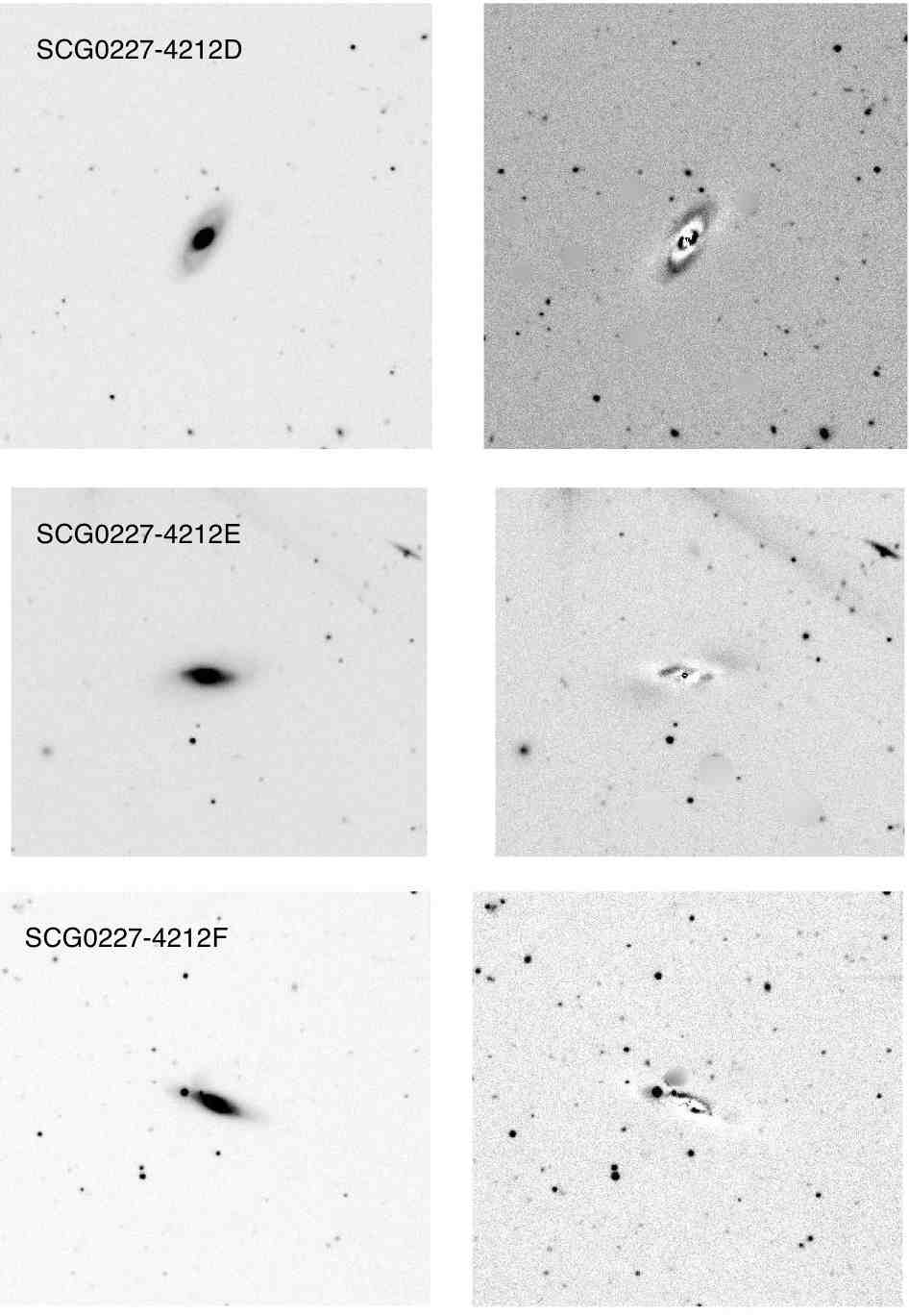}
\caption{ (continued) Optical images (left panel) and GALFIT residual images (right
panel) for SCG0227-4312, from galaxy {\it d} to {\it f} from top to bottom. 
Residual brightness for all galaxies is of the order of 3$\%$ of the
peak brightness. In all images N is up and E is left.}
\end{figure*}

\afterpage{\clearpage}

Table 3 shows the result from the GALFIT analysis on the member galaxies
of the six groups presented in the paper.

\setcounter{table}{2}
\onltab{3}{
\begin{table*}
%\begin{flushleft}
%\leavevmode
\caption{Results from the GALFIT analysis on the member galaxies of our SCGs.\newline
The symbols used here have the same meaning as in Peng 2002; the position angle is measured from N to E counterclockwise.
The magnitude is the total magnitude of the component, corrected for galactic extinction; errors on the measured 
magnitudes are of the order of 0.05 magnitudes and refer to the accuracy in the measurement of the zero-point
plus the sky subtraction and the statistical error from GALFIT.
Group SCG2159-3210 and galaxy SCG0122-3819{\it e} have not been observed by us in the optical; for photometric measurement 
the reader is referred to Hickson et al., 1982 and to NED.}
\label{scg_pho}
\begin{minipage}{\textwidth}
\begin{tabular}{lcccccccc}
\hline
Name of the galaxy & Functional form & Mag   & 
r$_{e,s}$\footnote{ r$_{e}$ is the effective radius of the Sersic or the deVaucouleurs's law, while r$_{s}$ is the scale length
of the exponential disk}  & n    & q    & PA        & c\footnote{Diskiness (negative) or boxiness (positive) paramenter. It has no 
real meaning for central sources, so it is not reported for these cases.} &Comments\footnote{Activity classification for HCG 90 is 
given in Coziol et al. (1998), while for all other groups, except SCG0141-3429, is from Coziol et al. (2000).}           \\ 
                   &                 & (R)   & (arcsec)        &      &      & (deg)     &            &                             \\ \hline   
SCG\,2353-6101A & Sersic          & 15.14 & 2.14            & 0.76 & 0.79 & -59.1     &  -0.01     & (R'\_1)SAB(rs)b, SFG         \\
                & Sersic          & 15.42 & 10.46           & 0.5  & 0.29 & -75.2     &    -       & central source, point-like  \\
                & Expdisk         & 13.26 & 10.21           &  -   & 0.71 & -70.8     &   -0.18    & strong, twisted spiral arms \\
SCG\,2353-6101B & Sersic          & 16.24 & 11.59           & 1.29 & 0.85 &  68.9     &  -0.45     & Sc{\it pec}, LINER          \\
                & Expdisk         & 13.52 & 14.33           &  -   & 0.34 &  87.8     &  -0.13     & very diffuse spiral arms    \\
SCG\,2353-6101C & Sersic          & 14.27 & 16.18           & 2.11 & 0.81 &  79.0     &   0.08     & S0                          \\
                & Sersic          & 14.8  & 1.4             & 0.64 & 0.98 &  83.6     &    -       & central source, point-like  \\
                & Expdisk         & 13.74 & 3.96            &  -   & 0.42 &  83.1     &   -0.16    &                             \\
                &                 &       &                 &      &      &           &            &                             \\
SCG\,0018-4854A & Sersic          & 13.68 & 6.20            & 1.99 & 1.00 &  80.3     &   -0.7     & Sa{\it ap}                 \\
                & Sersic          & 12.76 & 31.18           & 2.26 & 0.42 & -22.8     &   -0.28    &                             \\
                & Expdisk         & 13.13 & 4.94            &  -   & 0.62 & -32.7     &   -0.17    & extended tidal tail         \\
SCG\,0018-4854B & Expdisk         & 12.72 & 6.66            &  -   & 0.52 & -48.9     &   -0.05    & SB0/a{\it pec}, Sy2         \\
SCG\,0018-4854C & Expdisk         & 13.66 & 7.79            &  -   & 0.79 &  5.4      &   -0.49    & Irr, SFG                    \\
SCG\,0018-4854D & Sersic          & 15.90 & 2.01            & 1.00 & 0.92 & -17.0     &    0.21    & SBab{\it pec}, LINER       \\
                & Expdisk         & 14.01 & 5.91            &  -   & 0.75 & -21.0     &    0.35    &                             \\
SCG\,0018-4854E & Sersic          & 15.90 & 2.01            & 1.00 & 0.92 & -17.0     &    0.21    & SAB(r)c, SFG                \\
                & Expdisk         & 14.01 & 5.91            &  -   & 0.75 & -21.0     &    0.35    &                             \\
                &                 &       &                 &      &      &           &            &                             \\
SCG\,0122-3819A & Sersic          & 13.61 & 3.66            & 1.19 & 0.85 & -23.7     &    0.03    & SBa                         \\
                & Bar             & 14.10 & 42.3            &  -   & 0.77 & -30.0     &    0.01    & flat profile bar            \\
                & Expdisk         & 13.26 & 15.2            &  -   & 0.37 & -56.1     &   -0.29    &                             \\
SCG\,0122-3819B & DV              & 13.72 & 22.3            &  -   & 0.42 &  39.2     &   -0.15    & SB0, LLAGN                  \\
                & Bar             & 14.10 & 42.3            &  -   & 0.77 & -30.0     &    0.01    & flat profile bar            \\
                & Expdisk         & 14.60 & 4.12            &  -   & 0.43 &  43.1     &   -0.23    & strongly twisted spiral arms \\
SCG\,0122-3819C & Sersic          & 13.72 & 22.3            & 1.09 & 0.42 &  39.2     &   -0.15    & Sc, SFG                      \\
                & Expdisk         & 14.60 & 4.12            &  -   & 0.43 &  43.1     &   -0.23    & strongly twisted spiral arms \\
SCG\,0122-3819D & Sersic          & 15.11 & 3.11            & 1.00 & 0.66 & -79.6     &   -0.29    & S0B, exponential bar         \\            
                & Expdisk         & 14.53 & 4.77            &  -   & 0.99 & -78.9     &    0.17    &                              \\
                &                 &       &                 &      &      &           &            &                              \\
SCG\,0141-3429A & Sersic          & 14.12 & 1.97            & 0.79 & 0.61 & -58.8     &   -0.01    & S0$^{+}$                     \\
                & Expdisk         & 13.11 & 8.00            &  -   & 0.37 & -56.1     &   -0.29    &                              \\ 
SCG\,0141-3429B & Sersic          & 16.60 & 4.31            & 0.65 & 0.33 &  32.7     &   -0.59    & SAB(s)bc{\it pec}           \\
                & Sersic          & 13.98 & 24.44           & 0.5  & 0.27 &  49.5     &    0.02    &                              \\
SCG\,0141-3429C & Sersic          & 14.29 & 10.40           & 0.51 & 0.21 &  17.5     &   -0.13    & SB(rs)0, SFG                 \\
                & Expdisk         & 15.11 & 6.41            &  -   & 0.66 &  17.3     &    0.83    &                              \\
SCG\,0141-3429D & Sersic          & 15.83 & 1.34            & 0.76 & 0.88 &  88.3     &     -      & S0                            \\
                & Sersic          & 14.07 & 9.55            & 1.29 & 0.29 &  87.9     &   -0.20    &                              \\
SCG\,0141-3429E & Sersic          & 14.92 & 19.22           & 0.93 & 0.21 &  84.5     &   -0.40    & Sc {\it sp}, SFG             \\
                &                 &       &                 &      &      &           &            &                              \\        
SCG\,0227-4312A & Sersic          & 11.55 & 42.81           & 2.95 & 0.79 &  34.8     &    0.09    & S0                           \\
                & Sersic          & 13.97 & 2.59            & 0.97 & 0.92 &  31.9     &   -0.04    &                              \\
SCG\,0227-4312B & Sersic          & 14.42 & 5.11            & 1.37 & 0.66 & -15.7     &   -0.16    & (R)SAB(rs)b, EMLG            \\
                & Sersic          & 13.18 & 17.05           & 0.53 & 0.89 & -64.0     &    0.09    &                              \\
SCG\,0227-4312C & Sersic          & 14.76 & 10.97           & 0.50 & 0.31 & -16.1     &   -0.33    & Sbc                           \\
                & Sersic          & 14.20 & 3.07            & 0.90 & 0.79 & -40.8     &   -0.06    &                              \\
                & Expdisk         & 12.29 & 27.60           &  -   & 0.65 & -52.0     &   -0.12    &                              \\
SCG\,0227-4312D & Sersic          & 13.79 & 15.22           & 1.49 & 0.44 & -40.8     &   -0.27    & SBb, LLAGN                   \\
                & Sersic          & 14.42 & 3.54            & 1.00 & 1.00 &  8.8      &   -0.58    &                              \\
SCG\,0227-4312E & Sersic          & 14.87 & 16.37           & 0.68 & 0.25 &  89.9     &    0.73    & S0, SFG                      \\
                & Sersic          & 14.37 & 7.65            & 2.02 & 0.83 & -89.4     &    0.85    &                              \\
                & Expdisk         & 16.11 & 7.32            &  -   & 0.25 &  76.4     &    0.79    &                              \\
SCG\,0227-4312F & Sersic          & 15.40 & 3.67            & 1.05 & 1.00 &  68.9     &     -      & Sa                           \\
                & Sersic          & 15.12 & 13.44           & 1.00 & 0.31 &  58.2     &    0.20    &                              \\
                & Expdisk         & 14.34 & 11.33           &  -   & 0.24 &  71.7     &    0.19    &                              \\ \hline
\end{tabular}
\end{minipage}
\end{table*}
}

\afterpage{\clearpage}

\end{document}